\documentclass{aa}

\usepackage{graphicx}
\usepackage{txfonts}
\usepackage{makecell}
\usepackage{bm}
\usepackage{placeins}
\usepackage{float}

\usepackage{color,xcolor,xspace,pdflscape,dblfloatfix}

\definecolor{mhi}{rgb}{0.6,0.0,0.6}

\newcommand{\equ}[1]{Eq.~\ref{eq:#1}}

\newcommand{\fig}[1]{Fig.~\ref{fig:#1}}

\newcommand{\tab}[1]{Table~\ref{tab:#1}}

\newcommand{\sect}[1]{Sect.~\ref{sec:#1}}
\newcommand{\app}[1]{Appendix~\ref{app:#1}}

\newcommand{\lcdm}[0]{$\Lambda$CDM\xspace}

\ExplSyntaxOn
\cs_new:Npn \expandableinput #1
  { \use:c { @@input } { \file_full_name:n {#1} } }
\AddToHook{env/tabular/begin}
  { \cs_set_eq:NN \input \expandableinput }
\ExplSyntaxOff

\begin{document}

\title{Globular cluster orbital decay in dwarf galaxies with MOND and CDM: Impact of supernova feedback}
    
\titlerunning{Globular cluster decay in dwarf galaxies}
\authorrunning{M. B\'ilek, F. Combes, S. T. Nagesh, M. Hilker}

   \author{
   Michal B\'ilek \inst{1,2},
   Fran{\c c}oise Combes \inst{1},
   Srikanth T. Nagesh  \inst{3},
   Michael Hilker  \inst{4}
   }

   \institute{Observatoire de Paris, LERMA, Coll\`ege de France, CNRS, PSL University, Sorbonne University, F-75014, Paris\\
    \email{michal.bilek@obspm.fr}
    \and FZU – Institute of Physics of the Czech Academy of Sciences, Na Slovance
1999/2, Prague 182 21, Czech Republic
    \and
    Strasbourg University, CNRS, Observatoire astronomique de Strasbourg, F-67000 Strasbourg, France
    \and
    European Southern Observatory, Karl-Schwarzschild-Strasse 2, 85748 Garching bei M\"unchen, Germany
             }
   \date{Received: January 2024; accepted March 2024}

 \abstract{Dynamical friction works very differently for Newtonian gravity with dark matter and in
 modified Newtonian dynamics (MOND). While the absence of dark matter considerably reduces the friction in major
 galaxy mergers, analytic calculations indicate the opposite for very small perturbations, such as 
 globular clusters (GCs) sinking in dwarf galaxies. Here, we study the decay of GCs in isolated gas-rich dwarf 
 galaxies using simulations with the Phantom of Ramses code, which enables both the Newtonian and the QUMOND MOND gravity. We modeled the GCs as point masses, and we simulated the full hydrodynamics, with star formation and supernovae feedback.
We explored whether the fluctuations in gravitational potential caused by the supernovae can prevent GCs from sinking toward the nucleus. For GCs of typical mass or lighter, we find that this indeed works in both Newtonian and MOND simulations. The GC can even make a random {walk}. However, we find that supernovae cannot prevent massive GCs ($M\geq 4\times10^5\,M_\sun$) from sinking in MOND. The resulting object looks similar to a galaxy with an offset core, which embeds the sunk GC. The problem is much milder in the Newtonian simulations. This result thus favors Newtonian over QUMOND gravity, but we note that it relies on the correctness of the difficult modeling of baryonic feedback. We propose that the fluctuations in the gravitational potential could be responsible for the thickness of the stellar disks of dwarf galaxies and that strong supernova winds in modified gravity can transform dwarf galaxies into ultra-diffuse galaxies.
 }

   \keywords{galaxies: star clusters: general --- galaxies: dwarf --- galaxies: evolution --
   galaxies: kinematics and dynamics  --- cosmology: dark matter}
               
   \maketitle

\section{Introduction}
\label{sec:intro}
The missing mass problem at multi-scale from galaxies to large-scale structures in the Universe is one of the most stubborn problems in astrophysics. {Since its discovery in galaxy clusters \citep{Zwicky1937}, then in galaxies through rotation curves \citep{Rubin1980}, and  its confirmation through gravitational lenses and cosmic background radiation \citep[e.g.][]{Planck2016}, many candidates for a dark matter particle have been proposed.} Perhaps the most favored candidate for cold dark matter (CDM) has been neutralino, the most stable particle from supersymmetry \citep[e.g.][]{Wimp2018}. However, no supersymmetric particles have been discovered yet at the Large Hadron Collider of CERN; therefore, other candidates have been seriously considered, such as the axion or axion-like particles \citep[ALPs;][]{Hui2017}. 

Alternatives have been proposed in terms of modified gravity, and one of the most successful is modified Newtonian dynamics (MOND), which was designed 40 years ago by \citet{milg83b,milg83c,milg83a}.  
MOND has several possible forms and can also be modified inertia. In our subsequent analysis, we chose modified gravity in its special flavor: quasi-linear modified Newtonian dynamics (QUMOND).
There exist well-known problems of MOND in galaxy clusters  \citep[e.g.,][]{famaey12} and in the early universe, where the acoustic peaks of the cosmic microwave background (CMB) cannot be reproduced, as well as in the power spectrum of large-scale structures. However, these issues might be in the process of being solved \citep{Skordis2021}. The predictions from MOND have been tested many times and compared to those from the standard CDM model for galaxies in the hope of choosing {the model that better fits the observations. For example, the existence and frequency of bars \citep{tiret07}, galaxy interactions \citep{tiret08}, and dynamical friction \citep{kroupacjp,bil19b, bil21} have been compared.} 
Recently, \citet{freundlich22} reported a failure of MOND in the Coma cluster, where ultra-diffuse galaxies (UDGs) reveal kinematics in accord with MOND as isolated galaxies, while the external field effect (EFE) should exist for galaxies in a cluster and reduce their expected internal velocity, with respect to the isolated case. But the problem of MOND in clusters has been known for a long time, and it awaits a new physical phenomena, such as dark baryons and/or screening effects, due to the introduction of another characteristic scale \citep{gqumond} expected to solve both problems. 

{Another test for MOND  is based on the observation of globular clusters (GCs) in dwarf galaxies, where the dynamical friction should be high and drive the GCs quickly into the nucleus \citep{nipoti08}.} Dynamical friction is a phenomenon very different in MOND and CDM. In the deep MOND regime, for very light galaxies and very small perturbations, such as a GC in a dwarf, analytical formulations have shown that the friction could be higher in MOND than in CDM \citep{ciotti04}. However, for large perturbations, such as those in mergers of two Milky Way-type galaxies, dynamical friction is much reduced compared to Newtonian gravity with dark matter, as shown by \citet{tiret08} and \citet{combes14}. A simple analytical formula has been proposed for the deep MOND dynamical friction on small perturbers in \citet{sanchezsalcedo06} and tested through simulations by \citet{bil21}. Such analytic calculations indicate that GCs should sink from their currently observed positions into the centers of low-mass, low-surface-brightness galaxies on the timescale of one gigayear. This contradicts the observations of 10-gigayear-old GCs in these galaxies, unless fine-tuning is invoked.

 \citet{bil21} have considered the decay of single or multiple GCs in an isolated UDG in the deep MOND regime. Using the Phantom of Ramses code (\textsc{por}; \citealp{Lughausen_2015}), they simulated the sinking GCs modeled as point masses to check whether all GCs could sink to the center of the system. What happened is that the GC sinking stopped at a fraction of a kiloparsec due to core stalling \citep{hernandez98}. The apocenter of the GC first reduced rather quickly, and the decay slowed down when the apocenter reachd 0.5\,kpc. This phenomenon is due to the existence of a core in the UDG density profile, implying a harmonic potential, with particles not being able to absorb the energy of the GC. 

In these simulations, the interaction between GCs was not taken into account, neither the possible stripping nor the destruction of the sinking GC through the UDG tidal forces. Also, UDGs were simulated with collisionless particles without gas. In the present paper, we consider sinking GCs in isolated dwarfs, which are smaller and lighter gas-rich galaxies, with effective radii as low as 0.5\,kpc, and {the galaxies} rotate. The results of \citet{bil21} indicate that the dynamical friction timescale decreases with the mass of the galaxy, and thus we tested the survivability of GCs for the least massive galaxies that are known to have GCs. We show that the introduction of gas dynamics, star formation, and feedback can have a substantial effect on the dynamics of GCs. Medium- and low-mass GCs are totally prevented from sinking. The feedback, however, is not strong enough, at least in our simulations, to solve the problem of fast sinking for the massive GCs.

In Sect. \ref{sec:obs}, we compile observations of dwarf galaxies and their GC populations in order to define the initial conditions of the simulations. The technical details of the simulations are described in Sect.~\ref{sec:simul}. The results regarding the survivability of the GCs are displayed in Sect.~\ref{sec:results}, and in Sect. \ref{sec:other}, we discuss other interesting phenomena seen in the simulations. Sect.~\ref{sec:sum}  presents a summary of our work.

\begin{figure}[h!]
        \resizebox{\hsize}{!}{\includegraphics{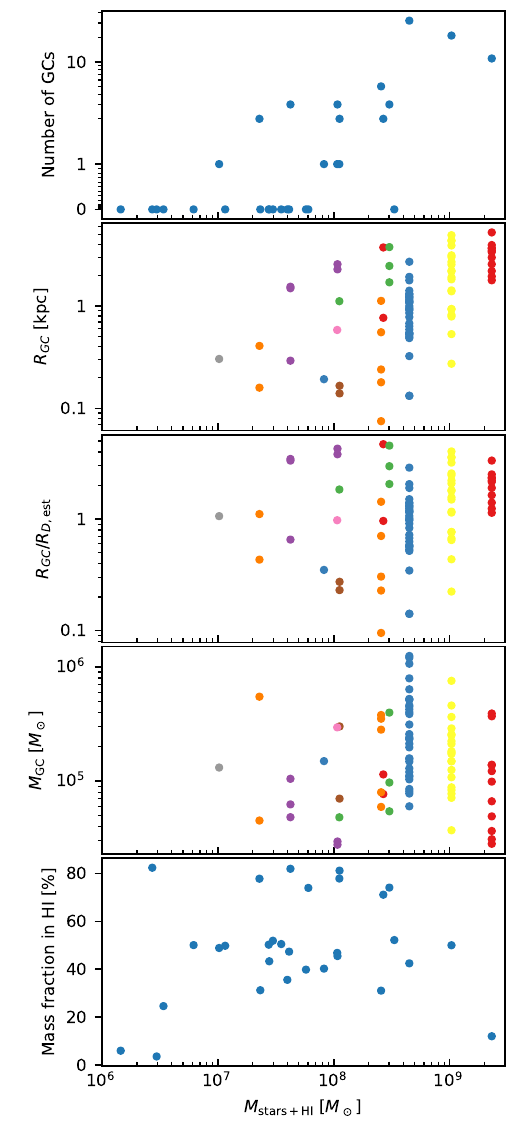}}
        \caption{Properties of observed isolated dwarf galaxies. From top to bottom: 1) Number of GCs in each galaxy, 2) projected galactocentric radius of each GC in each galaxy {($R_\mathrm{GC}$)} in the units of kiloparsecs, 3) same but in units of the scale lengths of the stellar disks of the galaxies {($R_\mathrm{D,est}$)} estimated from \equ{rdest}, 4) stellar mass of each GC {($M_\mathrm{GC}$)} in each galaxy, and 5) HI mass fraction in the total baryonic mass of each galaxy. The quantity on the horizontal axis is the total mass in stars and gas of each galaxy. The colors in the middle three tiles help distinguish between the GCs of different galaxies.   } 
        \label{fig:obsprops1}
\end{figure}

\begin{figure}[h!]
        \resizebox{\hsize}{!}{\includegraphics{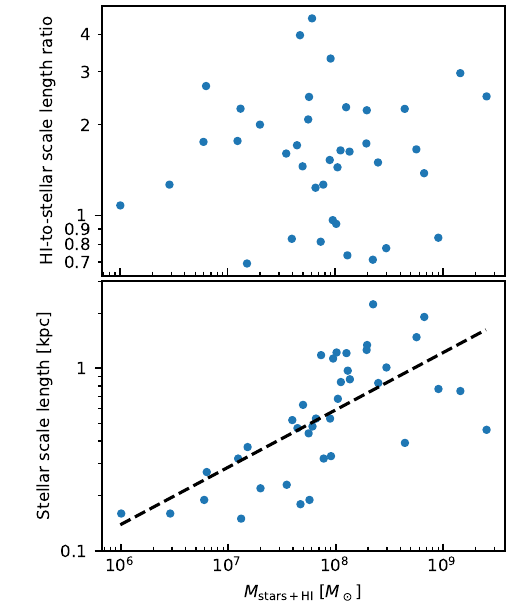}}
        \caption{Properties of observed isolated dwarf galaxies. From top to bottom: 1) Ratio of the scale lengths of the HI and stellar disks and 2) scale length of the stellar disk.  These quantities are plotted as functions of the total mass in stars and gas of the galaxy. The dashed line in the bottom panel shows the best linear fit given by \equ{rdest}.} 
        \label{fig:obsprops2}
\end{figure}

\section{Properties of observed isolated dwarf galaxies and of their GCs} \label{sec:obs}
In order to make simulations of relevant objects, we first  compiled information about the characteristics of the observed isolated dwarfs that are known to have GCs. It turns out that there are not many studies on GCs of isolated dwarfs. We took our data from the works of \citet{sharina05} and \citet{georgiev09}. The authors of each work were looking for GC candidates in \textit{Hubble} Space Telescope (HST) images of nearby dwarf galaxies ($2<D<11$\,Mpc with a few exceptions). 

Regarding the selection criteria, \citet{sharina05} and \citet{georgiev09} chose their GC candidates on the basis of their absolute magnitudes, colors, and structural photometric parameters since they appear spatially resolved in the HST images.
We included in our galaxy sample only the galaxies marked by the authors to be located in the field. {\citet{sharina05} estimated the contaminant fraction to be at most 10\% among their GC candidates. \citet{georgiev09} stated that their sample contains at most two contaminants per field. Nevertheless, the GCs of these nearby galaxies appear as resolved objects, which is one of the important selection criterion. Therefore, the contaminants are rather expected among the low-luminosity GC candidates.}

It was then necessary to obtain the characteristics of the galaxies. Their HI masses and distances were taken from the online database of the Catalog and Atlas of the Local Volume galaxies\footnote{\url{https://relay.sao.ru/lv/lvgdb/}}  \citep[the LV database hereafter, ][]{karachentsev19} in its form from April 1, 2022. Both \citet{sharina05} and \citet{georgiev09} provided absolute magnitudes of the galaxies in the $V$ band. We converted {the absolute magnitudes} to stellar masses, assuming the absolute magnitude of the Sun in $V$ of 4.83 \citep{binneymerrifield98} and the stellar mass-to-light ratio of one.
We neglected the mass of the possible molecular gas, helium, and metals. Because the distances of the galaxies are generally different in \citet{sharina05}, \citet{georgiev09}, and the LV database, we adopted for all galaxies the distances from the most recent source, namely, from the LV database. Accordingly, we corrected the stellar masses of the galaxies, {and} the projected galactocentric distances of the GCs. {The distance in the respective sources and in the LV database are given in \tab{distances}. The median deviation of the \citet{sharina05} distances  from the LV distances is 0.03\,dex (ca. 10\%), and for the \citet{georgiev09} data, this deviation is 0.006\,dex (ca. 1\%). }

The properties of the isolated dwarf galaxies are shown in \fig{obsprops1} as a function of the stellar plus HI mass of the galaxy. The figure shows that GCs are common in isolated dwarf galaxies more massive than $\sim10^8\,M_\sun$. The least massive galaxy that has a GC candidate has a mass of $1\times 10^{7}\,M_\sun$. The GCs have projected galactocentric distances of a few kiloparsecs. \citet{sharina05} and \citet{georgiev09} do not list the stellar disk scale lengths. However, after using \equ{rdest} derived below to estimate the stellar disk scale lengths, we found that the projected galactocentric radial distances of the GCs are comparable to stellar disk scale lengths. 
\citet{sharina05} nevertheless warns that their list of GC candidates can omit GCs that are far from their hosts because of the limited field of view of the HST camera.  In the fourth panel of \fig{obsprops1}, we show the estimated mass of the GC candidates, assuming a $V$-band mass-to-light ratio of 2.2 \citep{spitler09}. We note that this might be an overestimate because dwarf galaxies can also contain young GCs with lower mass-to-light ratios \citep{Pace2021}. The last panel tells us that isolated dwarfs are usually gas-rich objects. We found that the median HI mass fraction of the galaxies is 0.49. The median HI mass fraction of the galaxies with GCs is the same. 

It was also necessary to estimate the sizes of the galaxies because \citet{sharina05} and \citet{georgiev09} do not list them. We estimated them from the mass-size relation. The relation for galaxies of this type was extracted from the database of the LITTLE THINGS survey of gas-bearing dwarf irregular galaxies \citep{hunter12}. \citet{hunter21} listed for these galaxies the half-mass radii of the HI distribution from S\'ersic fits. We converted them to the estimates of scale-lengths by dividing them by 1.68, as if the HI gas were organized into exponential disks. Fitted exponential stellar disk scale-lengths were listed by \citet{hunter21} directly. The paper, however, does not list the stellar and HI masses of the galaxies. To estimate them, we converted the $V$ band magnitudes of the galaxies that they give to stellar masses, assuming the stellar mass-to-light ratio of one. We assumed, inspired by our main galaxy sample, an HI mass fraction of 0.5.

The results are shown in \fig{obsprops2}. The scale lengths of the HI and stellar disks differ, in median, by 0.2\,dex; that is, the HI disk size is typically 1.6 times the stellar scale length. The galaxies follow a well-defined mass-size relation. The best linear fit {of the logarithms of the baryonic mass and radius} reads
\begin{equation}
    \log_{10}R_\mathrm{D,*} = -2.74 + 0.314 \log_{10}(M_*+M_\mathrm{HI}).
    \label{eq:rdest}
\end{equation}
The stellar size-mass relation is rather independent of redshift for low-mass star-forming galaxies \citep{nedkova21}.
The axial ratio of HI disks of dwarf galaxies is about 0.6 \citep{roychowdhury10}.

To put the GCs of the considered dwarfs in a context of galaxies with known, well-sampled GC systems, we note that the absolute magnitudes of the GCs of a given galaxy generally follow a Gaussian distribution. While the width of the distribution can vary with the mass of the galaxy, the distribution usually peaks near the absolute $V$-magnitude of $-$7.5 \citep{rejkuba12}. Using the same assumptions as before, this corresponds to the mass of the GC of $2\times 10^5\,M_\sun$. This is close to the typical mass of the GCs in our sample.

\begin{figure}
        \resizebox{\hsize}{!}{\includegraphics{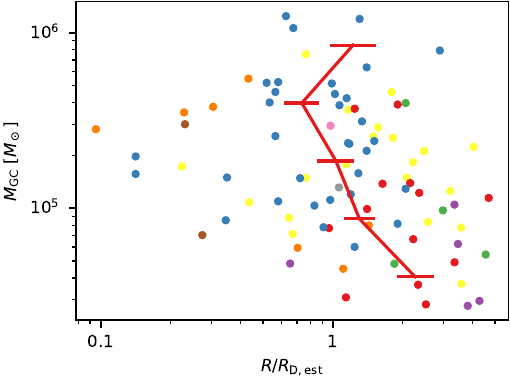}}
        \caption{Mass versus projected galactocentric radius for {the observed} GCs. The GCs have the same colors as in \fig{obsprops1}. 
        } 
        \label{fig:segreg}
\end{figure}

Figure~\ref{fig:segreg} shows the relation between the mass of the GC and its projected distance to the galaxy center. This distance is expressed in the units of the estimated stellar effective radius of each host galaxy.  The line with error bars in the figure indicates averages in bins and the uncertainties. If the dynamical friction strongly affects the GCs, then we expect the more massive GCs to be on average at lower galactocentric distances than the less massive GCs. Indeed, there is a tendency for such a correlation. {The Spearman test gives that the probability that there is no correlation is just 0.2\%.  }
 Our host galaxies hold up to 3\% (10\%) of their baryonic (stellar) mass in GCs.

\section{Numerical methods}\label{sec:simul}
We used the \textsc{por} code to run all the simulations. The code is a patch to the publicly available adaptive mesh refinement (AMR) grid-based code \textsc{ramses} \citep{Teyssier_2002}. The \textsc{por} patch was added to the 2015 version of \textsc{ramses} by \citet{Lughausen_2015}, and it implements the quasi linear formalism of MOND (QUMOND; \citealp{qumond}). The governing equation of QUMOND is

\begin{equation}
\nabla \cdot \bm{g} = \nabla \cdot \left( \nu \bm{g}_{_N} \right) \, ,
    \label{eq:poisson}
\end{equation}

\noindent where $\bm{g}$ and $\bm{g_N}$ are the true and Newtonian acceleration vectors. At each step, the Newtonian acceleration $\bm{g_N}$ is obtained using only the baryonic distribution, and subsequently, the interpolating function $\nu$ is determined at every step, which enables the code to compute the source term $\bm{g}$. The "simple" form of the interpolating function used in \textsc{por} is 

\begin{eqnarray}
    \nu ~=~ \frac{1}{2} + \sqrt{\frac{1}{4} + \frac{a_{_0}}{g_{_N}}} \, .
    \label{eq:Interpolationfunc}
\end{eqnarray}

\noindent It relates the MOND and Newtonian regimes \citep{famaey05}, and it has also been shown to work observationally \citep{gentile11, iocco15, banik18, Chae_2018}.

\subsection{Hydrodynamics, star formation, and feedback}
\label{sec:feedback}

As mentioned earlier, \textsc{por} only modifies the Poisson gravity solver from a Newtonian one to a MOND one, but it inherits the hydrodynamical solver and star formation and feedback recipes from the 2015 version of \textsc{ramses}. A second-order Godunov scheme with a Riemann solver for the conservative Euler equations are used \citep{Teyssier_2002}.

Star formation was modeled according to the Schmidt law (see, e.g., \citealp{shi11} for a discussion).  \citet{schmidt59} proposed that the stellar mass formed per unit time in unit volume, $\dot \rho_*$, is related to the local density of interstellar material, $\rho$, as a power law $\dot \rho_*\propto\rho^\nu$. As it was found observationally, the projected versions of these quantities are related as $\dot\Sigma_*\propto\Sigma^\nu$, where $\nu\approx 1.4$ \citep{kennicutt89}. This makes sense because, from the dynamical point of view in Newtonian gravity, we expect a relation of the type 
\begin{equation}
\dot \rho_* = \epsilon_*\frac{\rho}{t_\mathrm{ff}},
\label{eq:epsstar}
\end{equation}
where the free fall time is $t_\mathrm{ff} = \sqrt{3\pi/32G\rho}$ \citep[e.g.,][]{katz92}. In \textsc{PoR}, it is possible to prescribe star formation either by using \equ{epsstar} \citep{teyssier10,teyssier13} or the equivalent form  \citep{Rasera_2006, Dubois_2008}
\begin{equation}
\dot \rho_* = \frac{\rho}{t_\mathrm{dep}},
\label{eq:tstar}
\end{equation}
which we used in our study. Introducing $n$ as the number density of hydrogen atoms corresponding to $\rho$, the gas depletion timescale in \equ{tstar}, $t_\mathrm{dep}$, is calculated from the user-supplied parameters $t_*$ and $n_*$ as $t_\mathrm{dep} = t_*\left(\frac{n}{n_*}\right)^{-1/2}$. At the same time, the parameter $n_*$, with the dimension of the number density of hydrogen atoms, also plays the role of the threshold above which star formation is enabled. This is motivated by the observed fact that at low gas densities, the star formation rate is much lower than expected from extrapolating the Schmidt relation derived for denser gas. Both  $t_*$ and $n_*$  are theoretically poorly constrained, depend on the resolution of the simulation, and have to be tuned for the specific purpose. To derive the expression that \textsc{PoR} uses to model the star formation, {both sides of \equ{tstar} have to be multiplied by the volume of the computational cell $V_\mathrm{cell}$ and by the timestep $\mathrm{d}t$, and divided by the user-supplied mass of a star particle, $m_*$. In turn, the ideal number of stellar particles formed in a given time step would be}
\begin{equation}
\mathrm{d}N_* \equiv \frac{\dot \rho_*\, V_\mathrm{cell}\, \mathrm{d}t}{m_*} =  \frac{m_\mathrm{cell}}{m_*} \left(\frac{n_\mathrm{cell}}{n_*}\right)^{1/2}\frac{\mathrm{d}t}{t_*},
\label{eq:n}
\end{equation}
where $m_\mathrm{cell} = \rho_\mathrm{cell}V_\mathrm{cell}$ is the gas mass in the cell and $n_\mathrm{cell}$ is the number density of hydrogen atoms in the cell. The quantity $\mathrm{d}N_*$ is generally not and integer. Therefore, the actual number of stars formed in the given cell is drawn from a Poisson distribution that has the mean of $\mathrm{d}N_*$. In \textsc{PoR}, \equ{tstar} is activated by setting $t_*>0$; otherwise, \equ{epsstar} is used.

Supernova (SN) feedback plays a vital role in regulating star formation in a galaxy. Different types of feedback prescriptions available in \textsc{por} are described in detail in Sec. 2.2 of \citet{Nagesh_2023}, of which the intermediate feedback (Sec.2.2.1 of \citealp{Nagesh_2023}, originally implemented by \citealp{Dubois_2008}) is used in the present work. The intermediate feedback prescription allows the user to specify the fraction of energy from the SNe explosion to be injected back into the interstellar medium as kinetic energy, which is carried by a SN blast wave with a user-specified blast radius, $r_\mathrm{bubble}$, where velocity of the wave is computed using a local Sedov blast wave solution. Every time a stellar particle forms, metallicity, energy, and momentum are released into the respective cells. As far as the mass is concerned, whenever a stellar particle with mass $m_*$ is formed, $m_*(1+\eta_{sn})$ is removed from the gas cell, with $\eta_{sn}$ being a user specified fraction of mass of the new stellar particle that goes into the SNe. Subsequently,  $m_{\star} \eta_{\textrm{sn}}$ is injected back into the interstellar medium after a delay of $t_{sn}$ \citep{Dubois_2008}. The feedback parameters used are listed in Appendix~\ref{app:parameters}. 

Since its development, \textsc{por} has been applied on diverse scenarios using $N$-body-only polar ring galaxies \citep{lughausen13}, stellar streams \citep{thomas17}, the Local Group \citep{bil18}, and galaxies with shells \citep{Bilek_2022b}, as well as hydrodynamical simulations, the antennae pair of interacting galaxies \citep{renaud16}, disk stability of M33 in MOND \citep{Banik_2020_M33}, the UDG AGC 114905 with star formation \citep{Banik_fake_inclination_2022}, and the satellite plane \citep{Banik_2022_satellite_plane}. The code \textsc{por} has also been used to run a cosmological hydrodynamical simulation in the MONDian framework \citep{Wittenburg_2023}. \citet{Nagesh_2023} compared the effect of two types of feedback on galaxy models with stellar mass ranging between $10^7$ -- $10^{11}\,M_\odot$ and have found that the star formation rates and gas consumption timescale are in agreement with observations. 
Similarly, the effect of these feedback prescriptions on a monolithic cloud collapse of a rotating (non-rotating) gas cloud into disk (elliptical) galaxies in MOND have been investigated, too \citep{wittenburg20, Eappen_2022}. {Simulations of the formation of disk galaxies in MOND automatically lead to the correct scale lengths and exponential disks \citep{wittenburg20}.} {The monolithic collapse of non-rotating gas clouds can naturally explain the emergence of compact massive relic galaxies within a MOND-based universe \citep{eappen24}.} A user manual to set up $N$-body and hydrodynamical isolated disk galaxies in MOND using \textsc{por} \citep{Nagesh_2021}, as well as with \textsc{por} and other relevant packages, is available here.\footnote{\label{PoRbit}\url{https://bitbucket.org/SrikanthTN/bonnPoR/src/master/}}

\subsection{Generating initial conditions}

The Disk Initial Condition Environment (\textsc{dice}; \citet{Perret_2014}), adapted to MOND gravity by \citet{Banik_2020_M33}, was used to generate the initial conditions for {the} MOND simulations$^{\ref{PoRbit}}$. {The} details of the implementation of {the MOND gravity} in \textsc{dice} are discussed in \citet{Banik_2020_M33}. For the prescribed properties of the galaxy to be prepared, the code returns a list of positions, velocities, and masses of the initial stellar particles and a table with the rotation curve. Based on the rotation curve {provided by the MOND \textsc{dice} code}, {the} \textsc{por\_hydro} patch in {the} BonnPoR package$^{\ref{PoRbit}}$ calculates and generates the necessary distribution of {the} gas in the galaxy \citep{Banik_2022_satellite_plane}. We used the MAGI code \citep{magi} to generate the initial conditions for the Newtonian models. We used the code to generate three components in the galaxy: the stellar disk, the gas disk, and a dark NFW halo \citep{nfw}. The characteristic density and scale radius of the halo were chosen such that the galaxies follow the mean stellar-to-halo mass relation \citep{behroozi13}  and the halo mass-concentration relation \citep{diemer15}.  The stellar mass to generate these values was assumed to be half of the total initial baryonic mass of the galaxies because this is the typical value observed for our galaxies (\sect{obs}). The galaxies created by MAGI consist only of particles. In order to include gas, we calculated the rotation curve of the galaxy, removed the gas-disk particles, and let \textsc{por} deduce and generate the distribution of gas, as in the MOND case. {The stability of a halo created by MAGI is demonstrated in \app{nfwstab}.} {We modeled the GCs as point masses. In most of this work, the GCs had a mass of $10^5\,M_\sun$, which is about the typical mass of GCs in real dwarfs (\sect{obs}). In \sect{masses}, we explore how the results change for different GC masses.}

\subsection{Description of the galaxy models}
We were interested in whether GCs can survive in isolated dwarf galaxies without sinking to their centers for 10\,Gyr. This time was chosen because this is the typical age of GCs of the Milky Way \citep{Massari2023}. We initiated the simulated galaxies to be mostly gaseous because real galaxies were like this at the epoch of GC formation \citep{Tacconi2010}.

\begin{table*}[t!]
 \centering
\caption{Parameters of the basic MOND models used to generate the initial conditions in MOND DICE.}
\label{tab:mondmodels}    
\centering                              
\begin{tabular}{lllllllll}
\hline\hline                
Name & $M_\mathrm{bar}$        & $r_\mathrm{D}$ & $q$ & $f_\mathrm{gas}$ & $T$ & $Q$ & $N$  & $r_\mathrm{GC}$ \\
     & [$10^8\,M_\sun$] & [kpc] &     &           & [K] &   &  & [kpc] \\
\hline
GDw1e8  & 1   & 0.8 & 0.4 & 0.9 & 25000 & 0.9\tablefootmark{f}  & $10^5$ & 1   \\
GDw2e7  & 0.2 & 0.5 & 0.4 & 0.9 & 25000 & 0.9\tablefootmark{f}  & $10^5$ & 0.5 \\
GUDG1e8 & 1   & 2   & 0.5 & 0.9 & 25000 & 1.25\tablefootmark{m} & $10^5$ & 2   \\
GUDG2e7 & 0.2 & 2   & 0.5 & 0.9 & 21000 & 1.25\tablefootmark{m} & $10^5$ & 2   \\
\hline                                  
\end{tabular}
\tablefoot{ $M_\mathrm{bar}$: Total baryonic mass. $r_\mathrm{D}$: Disk scale length (the same for the stellar and  gas disk).  $q$: Axial ratio of the disk. $f_\mathrm{gas}$: Gas fraction. $T$: Gas temperature. $Q$: Toomre parameter (type:~ \tablefootmark{f}{fixed}, \tablefootmark{m}{minimal}). $N$: Number of particles in the stellar disk. $r_\mathrm{GC}$: Initial distance of the GC from the galaxy center.}
\end{table*}

\begin{table*}[t!]
 \centering
\caption{Parameters of the basic Newtonian models {used to generate the initial conditions in PoR and MAGI}.}
\label{tab:newtmodels}                                 
\begin{tabular}{llllllll}
\hline\hline                
Name & $M_\mathrm{bar}$ & $r_\mathrm{D}$ & $q$ & $f_\mathrm{gas}$ & $T$  & $\sigma_0$ & $Q_{R_D}$ \\ 

     & [$10^8\,M_\sun$] & [kpc] &     &           & [K] & [$\frac{\mathrm{kpc}}{\mathrm{Myr}}$]  & \\

\hline
GDw1e8N & 1   & 0.8 & 0.4 & 0.9 & 23000 & 5   & 1   \\  
GDw2e7N & 0.2 & 0.5 & 0.4 & 0.9 & 23000 & 0.3 & 1 \\ \hline\hline  

Name &  $N_*$  & $M_\mathrm{vir}$ & $r_s$ & $r_\mathrm{trunc}$ & $w_\mathrm{trunc}$   & $M_\mathrm{h}$   & $N_\mathrm{h}$ \\
     &[$10^6$] & [$10^9\,M_\sun$] & [kpc] & [kpc]              & [kpc] & [$10^9\,M_\sun$] & [$10^6$]       \\
\hline
GDw1e8N & 0.1 & 21   & 4.83 & 15 & 2 & 7.5 & 8 \\
GDw2e7N & 0.1 & 6.6  & 2.98 & 15 & 2 & 3.3 & 8 \\

\hline                                  
\end{tabular}
\tablefoot{$M_\mathrm{bar}$: Total baryonic mass. $r_\mathrm{D}$: Disk scale length (the same for the stellar and gas disk).  $q$: Axial ratio of the disk. $f_\mathrm{gas}$: Gas fraction. $T$: Gas temperature. $\sigma_0$ and $Q_{R_D}$: Special parameters of the MAGI code (see \citealp{magi}). $N_*$: Number of particles representing the stellar disk.  $M_\mathrm{vir}$: Virial mass of the NFW dark halo (if the halo were not truncated). $r_s$: Scale radius of the NFW dark halo. $r_\mathrm{trunc}$: Truncation radius of the halo. {$w_\mathrm{trunc}$: Width of the truncation.} $M_\mathrm{h}$: Actual mass of the halo (after applying the truncation).  $N_\mathrm{h}$: Number of particles representing the halo. }
\end{table*}

\begin{table}
\caption{Setup of the PoR code for the basic MOND models.}
\label{tab:code}
\centering
\begin{tabular}{ll}
\hline\hline                   
Parameter  & \makecell[l]{ Value (GDw1e8, GDw2e7, GUDG1e8, \\ \hspace{1em}  GUDG2e7)}                \\
\hline 
\texttt{levelmin} & 6,6,7,7\\
\texttt{levelmax} & 12,15,18,18 \\
\texttt{boxlen} & 150\,kpc \\
\texttt{mass\_sph}  & 1e-9 \\
\texttt{m\_refine} & 2e3 \\
\hline                                            
\end{tabular}
\end{table}

\begin{figure*}[]
        \centering
        \includegraphics[width=17cm]{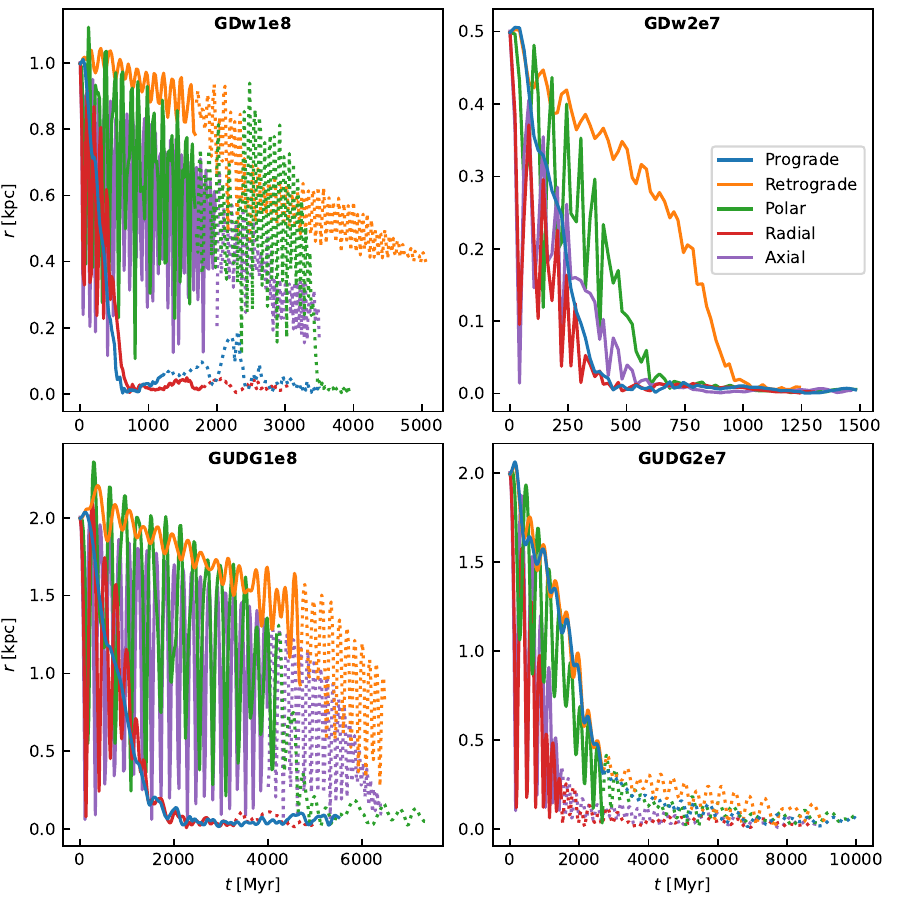}
     \caption{Evolution of the distance of the GC from the stellar barycenter of the galaxy for the MOND models without star formation. Each panel corresponds to the  galaxy model and each curve to a specific orbit of the GC. The dotted parts of the lines indicate unreliable parts: either the galaxy develops a bar that affects the motion of the GC, or (for GUDG2e7) the GC moves less than three resolution elements from the center.} 
        \label{fig:nosf}
\end{figure*}

\begin{figure}
        \resizebox{\hsize}{!}{\includegraphics{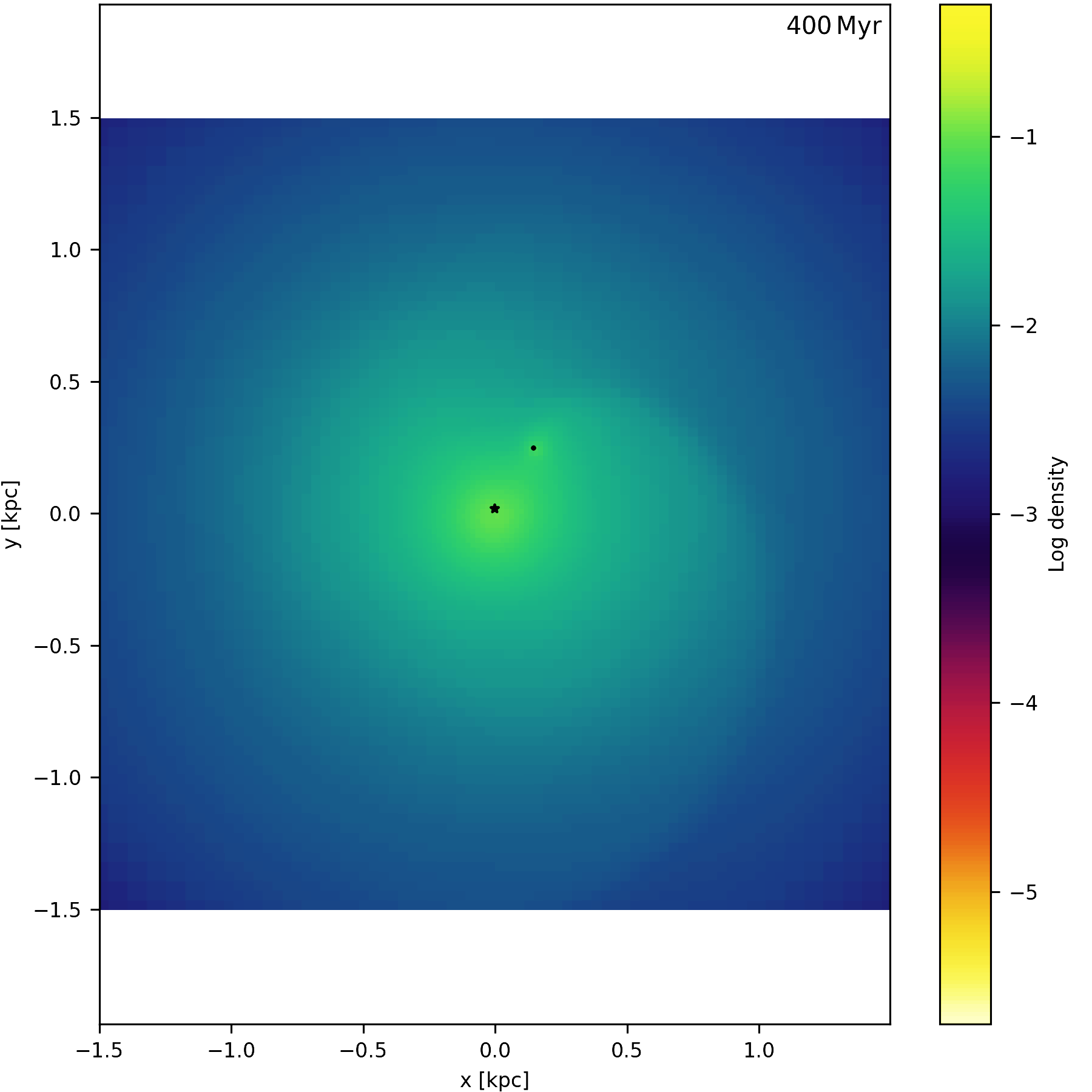}}
        \caption{Map of gas density in the midplane of the disk for the model GDw1e8 without star formation and with a GC on the prograde orbit. There is a density wave trailing behind the cluster. Both the GC and the material of the disk orbit the center of the galaxy counterclockwise.} 
        \label{fig:tail}
\end{figure}

\begin{figure}
        \resizebox{\hsize}{!}{\includegraphics{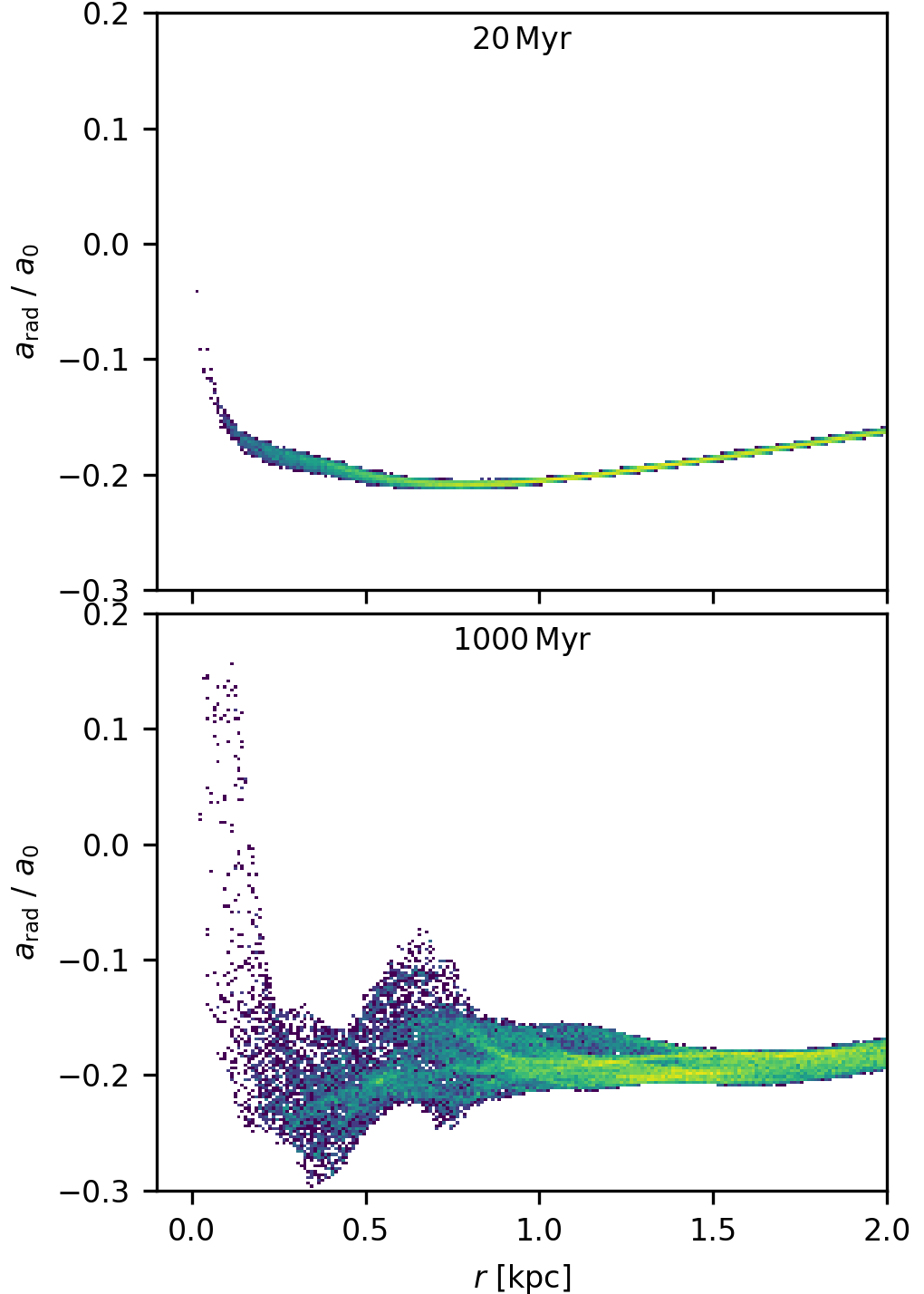}}
        \caption{Effect of the SN bubbles on the profile of the radial acceleration {$a_\mathrm{rad}$}. {A positive (negative) radial acceleration indicates that the vector points away from (toward) the galaxy center.} At the beginning of the simulation, before SNe appear, the profile is smooth (top). Later, the radial acceleration depends on the position within the galaxy (bottom).  The images are 2D histograms of individual stellar particles in the simulation GDw1e8\_e3. Only particles near the disk plane are shown here. The upper panel also shows that the gravitational potential of the galaxy is not harmonic (i.e., the acceleration is not $\propto r$).  } 
        \label{fig:accprof}
\end{figure}

\begin{figure*}[]
        \centering 

 \vspace{-2ex}
        \includegraphics[width=17cm]{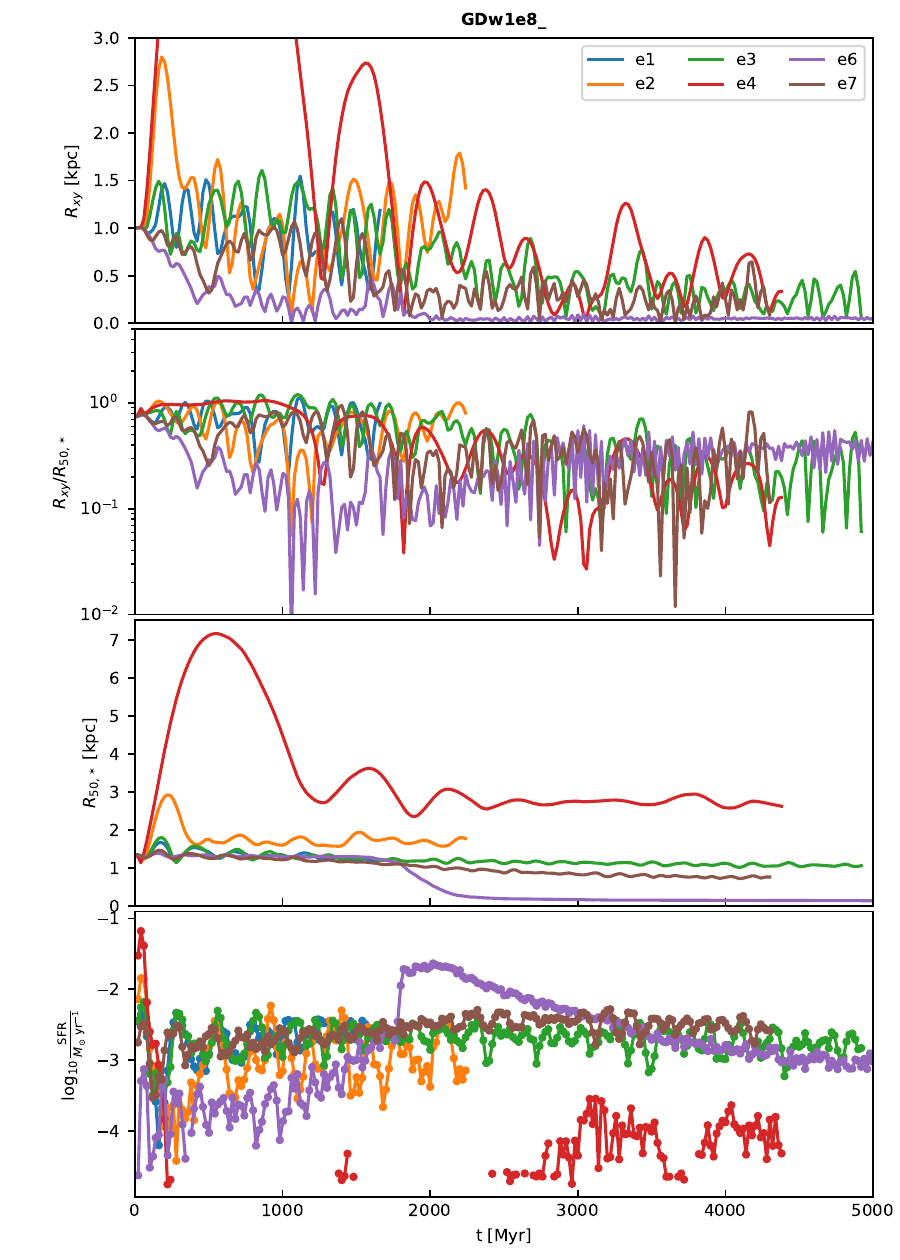}\vspace{-1ex}
     \caption{Different star formation variants of the model GDw1e8. The values of the star formation parameters are stated in \app{parameters}. From top to bottom, the panels show: 1) Evolution of the distance of the GC from the stellar barycenter of the galaxy projected to the disk plane {, $R_{xy}$,} in the units of kiloparsecs. 2) Same but in the units of stellar half-mass radius. 3) Evolution of the stellar half-mass radius {, $R_{50}$}. 4) Evolution of the star formation rate. {In the legend, we indicate just the suffix of the variant of the model indicated in the title of the plot (see the tables in \app{parameters}). }} 
        \label{fig:gdw1e8}
\end{figure*}

\begin{figure}[t!]
        \resizebox{\hsize}{!}{\includegraphics{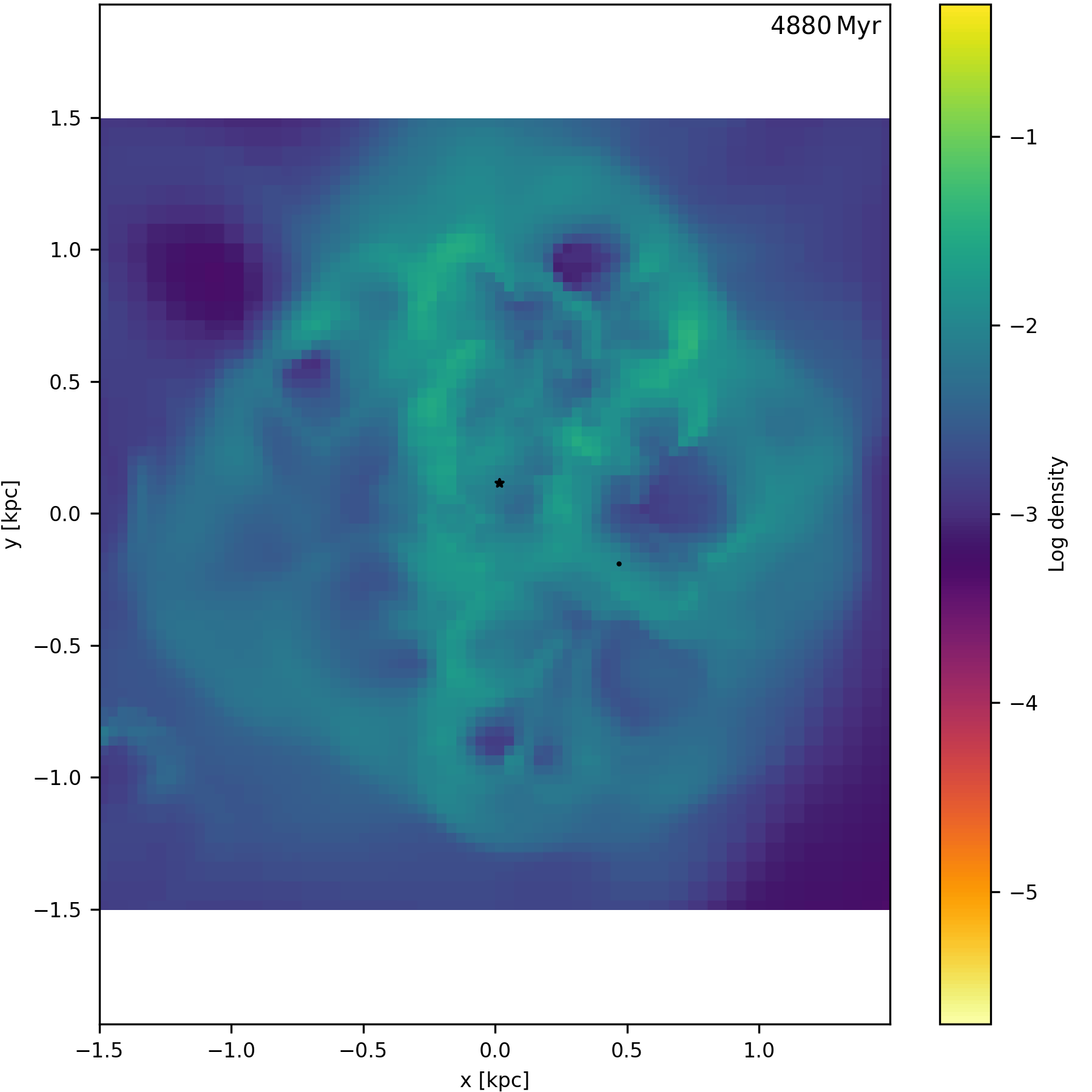}}
        \caption{Map of gas density in the midplane of the disk for the model GDw1e8\_e3. } 
        \label{fig:ex}
\end{figure}

\begin{figure*}[]
        \centering
        \includegraphics[width=17cm]{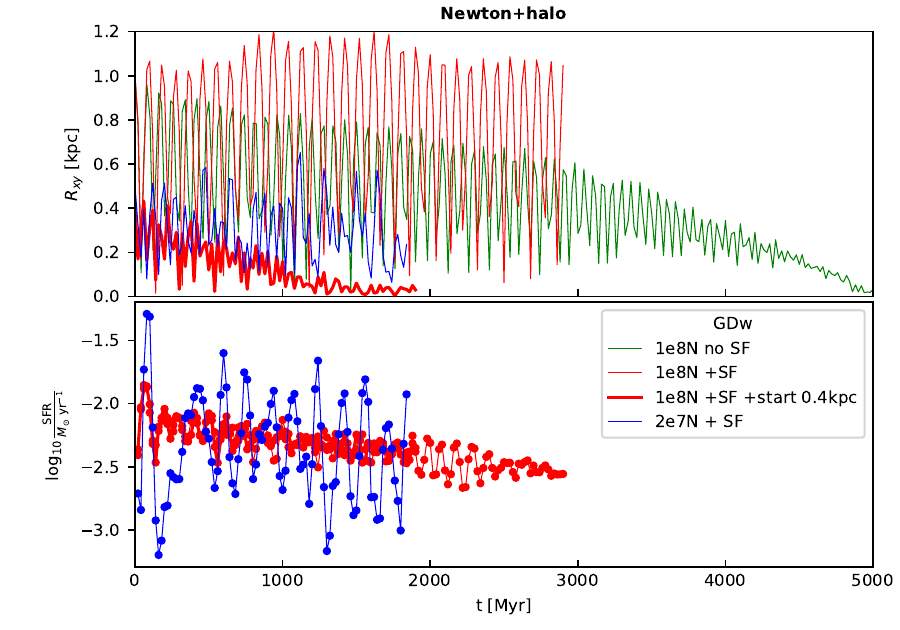}
     \caption{Newtonian models. The top panel shows the evolution of the distance of the GC from the particle barycenter of the galaxy. The bottom panel shows the evolution of the star formation rate.} 
        \label{fig:nfw}
\end{figure*}

\begin{figure}[t!]
        \resizebox{\hsize}{!}{\includegraphics{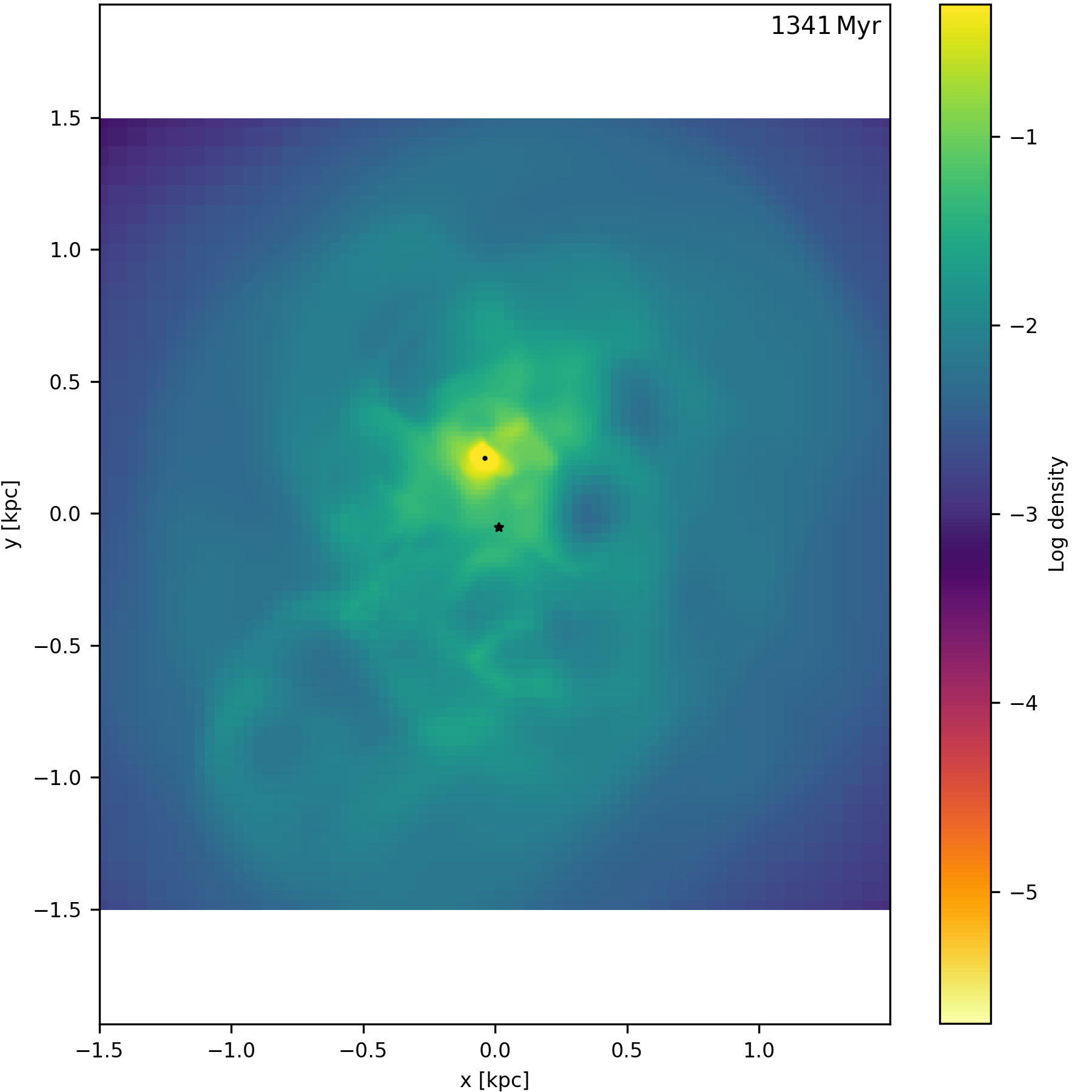}}
        \caption{Map of gas density in the midplane of the disk of the model GDw1e8\_e3. In this case, the mass of the GC was $4\times10^{5}\,M_\sun$ and moved on the retrograde orbit. The frame shows the final state of the simulation, when the GC, marked by the dot, settles in the offset central density peak of the galaxy. The stellar barycenter of the galaxy is nearly the center of the coordinate system and is marked by the asterisk. } 
        \label{fig:offset}
\end{figure}

We focused on four MOND models, GDw1e8, GDw2e7, GUDG1e8, and GUDG2e7, and for comparison, we considered two Newtonian models, GDw1e8N and GDw2e7N{ (see Tables~\ref{tab:mondmodels} and~\ref{tab:newtmodels})}. According to the scaling relations derived in \citet{bil21}, the sinking of GCs should be the fastest for the smallest dwarfs. The baryonic masses and radii of the models GDw1e8 and GDw1e8N were thus chosen according to the parameters of the observed dwarfs with a mass of $10^8\,M_\sun$ because this is the mass above which GCs in dwarfs start to be common (\sect{obs}). Similarly, the parameters of the models GDw2e7 and GDw2e7N were chosen to resemble those of the observed dwarfs with a mass of $2\times10^7\,M_\sun$ because this is the mass of some of the least massive galaxies that are known to have GCs. Actually, we could not have set a lower mass for the galaxy because the software we had at our disposal could not generate such galaxies and have them be stable. The initial scale lengths of the stellar and gas disks were the same. Mostly to explore the parameter space, we also considered the MOND models GUDG1e8 and GUDG2e7. They are more spatially extended than their counterparts GDw1e8, GDw2e7, and real isolated dwarf galaxies (\sect{obs}). They rather resemble observed isolated gas-rich UDGs \citep{Mancera-Pina2019}. These {galaxies} may host globular clusters \citep{Jones2023}.  The main characteristics of all the MOND and Newtonian models are presented in Tables~\ref{tab:mondmodels} and~\ref{tab:newtmodels}, respectively. Some experimentation and compromises were necessary in order to make the models relatively stable. This is why some of the parameters are not the same for all the models. In the MOND DICE,  we found, for example, that the Toomre  $Q$ parameter has very little influence on the stability of the galaxy.

Current-day simulations of galaxies do not have sufficient resolution to resolve all relevant processes. To model the gas physics, one would have to precisely model processes from the scale of astronomical units ($\sim 10^{11}\,$m) at which gas collapses into individual stars to the scale of kiloparsecs ($\sim 10^{19}\,$m) at which galactic tidal fields trigger the collapse of the gas clouds into stars. One thus resorts to approximate recipes for the subgrid physics. They are not only approximate, but they also contain parameters whose values are not known. Therefore, each of our models had several variants with different recipes for star formation and SN feedback in order to explore the possibilities. The variants of the models and their parameters can be found in the tables in \app{parameters}.

We limited the masses of the newly formed stellar particles to be at least 500\,$M_\sun$. This roughly corresponds to the minimum masses of OB associations, which can reach up to 10 000\,$M_\sun$ \citep{Massey1998}. This was necessary because RAMSES assumes that a certain fraction of mass of each of the newly formed particles explodes as in a SNe. All stars in the OB association that are destined to explode as SNe are assumed to die immediately after the stellar particle is created because the lifetimes of the very massive stars ($5\times10^{-3}$\,Myr) are shorter than the timestep of the simulation. Their explosion then occurs after the delay of 10 Myr mentioned in Sec. \ref{sec:feedback}.
If the masses of the new stellar particles in the simulation were too low, the SN feedback would be unrealistically spread in space and time. 

Unless stated otherwise, we always started the simulations with GCs at the apocenter and at the distance of one scale length of the galaxy. This roughly agrees with observed positions (\sect{obs}). In reality, the GCs could have been formed at larger distances. For the purposes of the current paper, where we are interested in whether the SN feedback can prevent GC sinking, this assumption seems appropriate.  Moreover, the fact that the distances of all the observed GCs are on the order of the stellar scale length of the galaxy indicates that it has always {been} the case; otherwise, we would encounter a fine-tuning problem. Moreover, as we show below, in the absence of star formation, the sinking times depend dramatically on the mass of the clusters, while only a mild correlation is observed between the masses and galactocentric radii (\fig{segreg}). This would make the fine-tuning problem even worse.

We modeled the GCs by point masses. We could not model them as resolved objects for the following reason. To model them as such, the simulation would have to have a spatial resolution of a fraction of a parsec so that we could model the internal dynamics of the GCs. At such a high spatial resolution, the mass in each gas cell would be so low that it would not cross our limit of $500\,M_\sun$ for the formation of new stellar particle, and no new stellar particles would be formed. 

The barycenter of the galaxy was always set as the origin of a right-handed Cartesian coordinate system {at the begining of the simulation}. The spin vector of the galaxy pointed in the $z$ direction, and the midplane of the disk was in the $x-y$ plane.

\section{Results}\label{sec:results}
\subsection{MOND models without star formation}
We first explored whether the GCs can survive in the galaxies without sinking for 10\,Gyr if star formation is disabled. In each of the four MOND models, we considered five orbits of the GC: 1) the GC moves on a circular orbit in the plane of the disk, corotating with it and starting at the $x$-axis (prograde orbit hereafter); 2) the same but the GC moves against the direction of the rotation of the disk (retrograde orbit hereafter); 3) the GC is initiated at the $x$-axis and is assigned the circular velocity in the direction of the $z$-axis (polar orbit hereafter); 4) the GC is dropped with a zero velocity with respect to the galaxy center from the $x$-axis (radial orbit hereafter); 5) the GC is dropped with a zero velocity with respect to the galaxy center from the $z$ axis (axial orbit hereafter). In all cases, the initial distance of the GC from the galaxy center was about one scale length of the disk (see \tab{mondmodels} for exact values). The GC in our simulations had a mass of $10^5\,M_\sun$. This value is roughly the typical mass of a GC (\sect{obs}).   

After some time, strong bars formed in most of the models. This time depended on the type of the orbit of the GC. The bars then affected the motions of the GCs substantially. The bar could form either before or after the sinking of the GC was completed.

The results are presented in \fig{nosf}. The dotted parts of the lines indicate the period when the galaxy has a bar or, for GUDG2e7 (which does not develop a bar), the period when the GC {moves} less than three resolution elements from the galaxy center. In all models, the GC on the prograde orbit sinks in a few gigayears.\footnote{A movie of the GC orbiting on the initially prograde orbit in the model GUDG1e8 can be seen at \url{https://share.obspm.fr/s/ncYbazAwBBPPeeB}} In contrast to the isolated gas-free UDGs explored in \citep{bil21}, there is no core stalling phase. The curve of the orbital decay of the GCs in the model GUDG2e7 show a slowdown of the decay once the GC approaches the core of the galaxy. We note, however, that during this period, the GC is separated just by about two resolution elements such that the sinking is not modeled properly. If the GC is put on the radial orbit, it sinks in almost the same amount of time, albeit mostly somewhat later. Snapshots from the simulations show that the GC feels the drag of the surrounding gas such that the GC starts corotating with the galaxy. The retrograde orbit is the least prone to orbital decay. In the small dwarf  GDw2e7, the sinking of the GCs is the fastest. For all the explored orbits, the sinking takes at most 1\,Gyr.  In all simulations, images show trailing waves behind the GCs (see the example in \fig{tail}). 

In \citet{bil21}, GCs of UDGs were showing core stalling because of the harmonic potential in the center of the galaxies. Here, we did not observe the stalling. Indeed, the upper panel of \fig{accprof} shows that the gravitational potential is not harmonic (the acceleration of a harmonic potential is proportional to the galactocentric radius). This plot was constructed from particles in the simulation GDw2e8\_e3  before the onset of SN explosions. Only the particles within 10\degr from the disk plane were used.

We explored whether the presence of two GCs can affect the survivability of at least one of them. The idea behind this was that one of the GCs would serve as a reservoir of energy for the other. One of the GCs of the model GUDG1e8 (we call it GC1) was placed on the prograde orbit with the initial radius of 2\,kpc. The second cluster, GC2, had the same mass and orbital plane but was on a retrograde circular orbit with a radius of 1\,kpc. We tried four different initial positions for GC2, offset from each other by ninety degrees. It turned out that GC2 does not affect the sinking time of GC1 appreciably. The reason is that the region in which the gravitational field of a GC dominates over the gravitational field of the galaxy is too small, and therefore the GCs influence each other too rarely to have an effect.

\subsection{MOND models with star formation}
For the simulations without star formation, we found that the orbital decay is usually the strongest if the GC is on the prograde orbit. For the simulations with star formation, we therefore explored only GCs initialized on the prograde orbit to see whether the SN explosion can prevent GCs from sinking even in this least favorable case.

Figures~\ref{fig:gdw1e8}-\ref{fig:gudg2e7} show how the models evolve.\footnote{A movie of the GC orbiting on the initially prograde orbit in the model GUDG1e8\_e1 can be seen at \url{https://share.obspm.fr/s/4pg2wdPdi3jrbKT}} The SN explosions can affect the trajectories of the GC substantially. Depending on how we set the baryonic physics parameters, the GC makes either larger or smaller oscillations in the galaxy. However, as long as there are SNe exploding (seen as a non-zero star formation rate in the figures), the GC does not sink completely. The SN explosions cause fluctuations in the density of the gas (\fig{ex}). These, in turn, give rise to fluctuations in the gravitational potential (bottom panel of \fig{accprof}). The GC experiences pushes from random directions, which prevent it from staying at any particular place in the galaxy. While the GCs initially move on prograde orbits, the random pushes eventually cause them to make random walks.

However, the SN feedback must have the right strength. If it is too weak, the GC stays very close to the center of the galaxy. This contradicts observations, where we see that GCs are located at around one scale length of the stellar disk (\sect{obs}). On the other hand, if the feedback is too strong, the galaxy explodes -- the gas is expelled, and since the gravitational field decreases, the stars spread as well. In MOND, for isolated objects, the escape velocity is infinite, so the galaxy cannot dissolve completely. It just becomes large in size and does not follow the observed mass-size relation (\sect{obs}). It is possible that some UDGs were formed from dwarf galaxies by such intensive starbursts (\sect{other}). 

The survival of the GC is not possible in the model GDw2e7. This is the model in which the sinking was the fastest in the simulations without star formation. We were not able to find star formation parameters that would lead to sustained star formation. The GC sunk shortly after the SNe had ceased. After the first SNe exploded, gas was heated up and diluted such that it stopped forming any further stars. It did not cool enough even after several gigayears. We suspect that this could be due to insufficient treatment of gas cooling in RAMSES. The code determines the cooling rate of a computational cell only on the basis of its temperature, density, and metallicity. There are no parameters to tune it. However, in a real galaxy, gas has a fractal structure \citep{Elmegreen1996}. There are compact dense cores of the size of a few astronomical units \citep{Pfenniger1994}, much smaller than the cells in simulations, in which cooling is much more efficient. Therefore it is not sufficient to consider just the average gas properties in the computational cell. In addition, {from the old stellar population the explosions of the type Ia SNe occur even without any local gas concentration;} however, they were not taken into account in the present simulations. The impact of SNe on the neighboring gas, and therefore the gravitational potential, might be underestimated. In any case, this example underlines the finding that SNe are essential for the survival of GCs in isolated dwarfs.

At first glance, it might appear surprising that SNe affect the trajectory of a GC. The SN forms a bubble that is a spherically symmetric structure. Outside of a sphere enclosing such a bubble, the gravitational field does not depend on whether gas inside the sphere is distributed homogeneously or if it is concentrated toward the surface of the sphere. Similarly, the gravitational force inside this spherical shell is zero. However, one needs to take into account that the galaxy is a disk, and the SN bubble forms a hollow ring with an increased density near its border. It is known, at least for Newtonian gravity, that the gravitational field inside a ring points outside the center of the ring. When viewing the simulation movies, it really looks as though the GC is pushed away from the centers of the SN bubbles. Moreover, if two or more SN bubbles overlap, as they often do in our simulations, the density disturbances are then not spherically symmetric (\fig{ex}).

It is interesting that the galaxies in our simulation have relatively low star formation rates. If the galaxies are to convert 50\% of their gas mass into stars in 10\,Gyr to reach the current observed gas fraction, then the average star formation rate must be, in decadic logarithm, -2.3 (-3.0) {$M_\sun\,$yr$^{-1}$} for the $1\times10^{8}\,M_\sun$ ($2\times10^{7}\,M_\sun$) galaxy. It turned out that achieving such a high star formation rate is difficult. The models with the highest star formation rates have star formation parameters that are quite different from the fiducial values (\sect{feedback}). While we could manage an intensive starburst right after the star of the simulation, the resulting SNe diluted the gas such that the subsequent star formation was strongly decreased. We were able to achieve higher star formation rates easily in the Newtonian simulations, as the massive dark halos prevent gas from being spread. Nevertheless, the consequence of this disparity between the observed and simulated star formation indicates that in the real Universe, there are more SNe than in our simulation. Therefore, the real effect of SNe on the motion of GCs and stars must be larger than in the simulations.

\subsection{Newtonian simulations}
We were interested in whether the problem of fast sinking of GCs is encountered in Newtonian gravity with CDM. In the CDM scenario, dwarf galaxies are dominated in mass by dark matter, even in the very center. Although the NFW profile predicts a DM cusp in the center, it is possible that it is mitigated by the star formation feedback, although the weak level of star formation might not be able to transform the cusp into a core. 

In the model GDw1e8N, we put the GC on a radial orbit with an initial distance of 1\,kpc. We expected the strongest dynamical friction when the GC goes through the center of the halo, where it has an infinite density \citep{nfw}. The result is shown in \fig{nfw} by the green curve. The GC sinks in 5\,Gyr. This indicates GCs have the potential to sink in the centers of their hosts even in the \lcdm cosmology.   

Next, we were interested in whether SNe can prevent such a cluster from sinking. The star formation parameters are in \tab{varnewt}. The simulations turned out to run much slower than the MOND simulations. This is why we continued the simulation only until ca. 3\,Gyr and had to adjust the standard star formation parameters (\sect{feedback}) in order to decrease the star formation rate. The result is shown by the thin red curve in \fig{nfw}. Unlike in the case without star formation, there is no hint of sinking within the simulation time, meaning that the SNe really prevent the GC from sinking. We then ran another simulation, placing the GC at an initial distance of just 0.4\,kpc from the galaxy center while keeping the star formation parameters the same. It again had a zero relative velocity with respect to the galaxy. The result is shown in \fig{nfw} by the thick red curve. In this case, the GC sinks in ca. 1.5\,Gyr. We repeated the exercise for the model GDw2e7N. The GC started from 0.5\,kpc. This time, we used the standard star formation parameters. The GC did not show any hints of sinking for the whole simulation time until ca. 2\,Gyr. 

{We conclude from these simulations that SN explosions can but does not always prevent GCs} from sinking in the context of \lcdm cosmology. The initial position and velocity of the GC is an important factor. 

It is worth noting that while we explored just three situations, the GCs approximately kept their initial apocentric distances in two of them. In most MOND simulations, the apocentric distance oscillates either around a larger or lower value. Moreover, unlike the MOND simulations, the star formation rate remained close to or higher than the value required by observations without any ad hoc tuning.

\subsection{Different masses of the GCs}
\label{sec:masses}
We also briefly explored whether our conclusions for the MOND models would change if we changed the mass of the GCs. According to the Chandrasekhar formula for dynamical friction and the results of the MOND simulations of \citet{bil21}, the sinking time should decrease when increasing the mass of the GC. 

Indeed, when the GC in the model GUDG1e8 was decreased to $10^4\,M_\sun$ on the prograde orbit, it did not have a problem surviving in the galaxy for 10\,Gyr, even without star formation. The galaxy, however, developed a strong bar in the simulation time of about 4\,Gyr, which influenced the GC substantially.

We then explored the effect of increasing the mass of the GC to $10^6\,M_\sun$. This is the maximum we encountered in the sample of the observed isolated dwarf galaxies (\sect{obs}), even if only in one galaxy. We entered this GC in the model GDw1e8, first on the prograde orbit, without star formation enabled.\footnote{A movie can be seen at \url{https://share.obspm.fr/s/G8c8nZ7mmKqK9YJ}} The sinking progressed differently than usual. The GC first accumulated a massive cocoon of gas and stars around itself, which effectively made its mass even larger. After finishing a single revolution around the galaxy barycenter, the GC captured the central density peak of the galaxy. This large overdensity kept orbiting the galaxy pericenter without approaching the barycenter much more. If observed, the galaxy would be classified as having lopsided outer parts of the disk, and the GC would be right in the middle of the galaxy center. This process took just about 300\,Myr.

If the $10^6\,M_\sun$ GC is put on a retrograde orbit, the situation is different.\footnote{A movie can be seen at \url{https://share.obspm.fr/s/WbEo7rgexAgdE3A}} Instead of capturing gas, the GC induced a V-shaped wave trailing behind it -- like the one that can be seen on the water behind a ship. The sinking was slower than in the prograde case. Before the GC reached the center of the galaxy, the wave formed an overdensity near the galaxy center, orbiting it in the direction of the disk rotation. Once the GC sunk close enough to this overdensity, the GC suddenly switched the direction of its orbit and stuck with the central density peak of the galaxy. Again, observationally, we would classify the object as a lopsided galaxy with a GC in its central density peak. Reaching this state took just about 700\,Myr. 

We then enabled star formation. The GC was put on the prograde orbit with an initial radius of 1\,kpc.\footnote{A movie can be seen at \url{https://share.obspm.fr/s/qwg9KyR4MLtGsTo}} The feedback parameters were set as in the simulation GDw1e8\_e3 because in that simulation, the GC was kept at large galactocentric distances the easiest when it had a mass of $10^5\,M_\sun$. In spite of that setup, with the massive GC, the sinking went virtually the same as without star formation. The SNe were too weak to affect the process substantially. The GC harbored in the center of the central density peak in  300\,Myr.   Exchanging four random stellar particles of the galaxy by GCs of this mass did not change the outcome much. All four GCs and the galaxy density peak merged into a single off-centered object in 540\,Myr. Then, back with a single GC in the simulation, we reduced the mass of the GC to $4\times10^5\,M_\sun$. This is more appropriate for a galaxy of this mass (\fig{obsprops1}). The GC {merged} with the central density peak in 500\,Myr. When this GC was put on the retrograde orbit, which is usually the orbit that is the least prone to sinking, the GC merged with the density peak in 1300\,Myr, as shown in \fig{offset}. Just as without star formation, 20\,Myr before merging, the GC suddenly reversed the direction of its motion. {It stuck with the stellar and gas density peaks, being offset with respect to the galaxy stellar barycenter.} {We also put the GC of the mass of $10^6\,M_\sun$ on a circular retrograde orbit inclined by  $45\degr$ with respect to the galactic disk. In this case, the GC captured the central density peak after 540\,Myr.} In total, in our MOND simulations, the SNe cannot prevent massive GCs from sinking into the center of the galaxy. More precisely, the central density peak of the galaxy and the GC stick together.

We then explored whether the situation is different in the Newtonian case. A GC with a mass of $4\times10^5\,M_\sun$ was entered in the model GDw1e8N with star formation and initiated to move on a retrograde orbit with a radius of  1\,kpc. Due to the slow progress of the simulation, we ran it only for 1800\,Myr so that we could compare it with the MOND case. The sinking was quite mild, just to 0.82\,kpc. Indeed, unless our simulations are wrong, Newtonian gravity with NFW dark halos can explain the observations of massive GCs in dwarf galaxies more easily than MOND.

\section{Other interesting phenomena seen in the  MOND simulations}
\label{sec:other}

The SNe bubbles randomly push not only  GCs but also stars, increasing their velocity dispersion. We saw in some of our MOND simulations that strong SN winds can disperse the gas, which holds most of the baryonic mass of the galaxy. Once the gas is expelled from the center of the galaxy, the gravitational field becomes weaker in this region such that the stellar component of the galaxy grows in radius too. This can be seen most prominently in the model GDw1e8\_e4, where the stellar half mass radius grew from  1\,kpc to 3\,kpc. Such a mechanism could be a formation channel of UDGs, at least for their small-mass end. Other formation mechanisms of UDGs proposed in the literature for Newtonian gravity include the scenario of gas outflows associated with star formation in several burst episodes within dwarf-sized halos \citet{DiCintio2017}, which is similar to ours. Also, tidal interactions and ram-pressure stripping have been invoked \citep{Carleton2019}.

In addition, the random pushing of stars by SNe bubbles makes the stellar disks thicker. This phenomenon plays a role in both stars and gas and particularly in the smallest mass galaxies. In agreement with this, stellar and gas disks of dwarf galaxies are relatively thick compared to more massive galaxies \citep[e.g.,][]{roychowdhury10,roychowdhury13}.

\section{Summary and conclusions}\label{sec:sum}

Analytic arguments indicate that GCs of low-mass, low-surface-brightness galaxies experience strong dynamical friction in MOND. It should cause the GC to sink in the centers of their host galaxies on timescales on the order of 1 Gyr. Observations of 10-gigayear-old GCs in such galaxies, however, have raised doubts about the validity of MOND -- or the correctness of the analytic calculations. Indeed, the simulations of GCs of isolated gasless UDGs by  \citet{bil21} showed that the GCs sink much slower than it initially appeared because of the core-stalling mechanism that was not taken into account in the analytic calculations. In the present work, we focused on GCs in isolated dwarf  galaxies and UDGs with MOND (more precisely QUMOND) and Newtonian gravity. Our main aim was to see whether SN explosions can prevent GCs from sinking.

We first explored the observational properties of isolated dwarf galaxies and their GCs on the basis of literature data (we are not aware of any observational studies of GCs of isolated UDGs; see \sect{obs}). Our main findings are the following: 1) Isolated dwarfs are usually gas-rich objects. The median gas fraction is 50\%. Given that the SNe only directly affect gas, this observation justifies the expectation that SNe can cause fluctuations of the gravitational potential and, as a consequence, dynamically heat the orbits of the GCs.  2) The least massive dwarfs do not contain GCs. The least massive dwarf in our sample has a mass of $10^7\,M_\sun$. Only dwarfs with masses over about $10^8\,M_\sun$ usually have GCs. This defined the galaxies of interest for our study. 3) The distribution of GCs does not seem to differ much from the distribution of the stars of the host galaxy. This indicates that they roughly formed together and are subject to similar forces; otherwise, we would run into a fine-tuning problem. 4) More massive GCs tend to be located closer to the center of the galaxy, with the possible exception of the most massive GCs (mass over about $5\times10^5\,M_\sun$), in agreement with the expectation that more massive GCs are more affected by dynamical friction. 

In the simulations, we initiated the galaxies to follow roughly the observed mass-size relations. They were initiated to be 90\% gaseous in order to mimic the conditions at the birth of the GCs 10\,Gyr ago. In most of the simulations we performed, the GC had a mass of $10^5\,M_\sun$, which is roughly the typical observed mass. The GC was modeled by a point mass. The GCs always started at the apocenter, at about one scale length of the disk.

We started with MOND simulations without star formation. There were four galaxy models with more and less massive dwarfs and more and less massive UDGs. Regardless of the direction of the initial velocity vector, the GCs sunk in the centers of their host galaxies on the timescale of 1 Gyr. Unlike in the simulations of GCs of UDGs \citep{bil21}, no core stalling occurred. The sinking was usually the fastest for the radial and prograde orbit, and it was the slowest for the retrograde orbit. However, in some cases it was not possible to study the orbital decay in its completeness because the galaxies became unstable and started reorganizing themselves. This process was halted or modified when the SN feedback was included later. Nevertheless, this demonstrated that if a bar occurs in the galaxy, it can push the GC and extend its lifetime.  

We then introduced star formation and SN feedback. Because star formation and SN explosions can be modeled only approximately in current simulations, we explored various subgrid parameters. The GC was always initiated on the prograde orbit, which is the most prone to sinking. We indeed found that SNe can prevent the $10^5\,M_\sun$ GCs from sinking. The GC can forget the direction of its initial orbital motion and make a random walk. However, it was quite difficult, or impossible, to choose the baryonic parameters such that the galaxies form an appreciable number of stars. In the low-mass dwarf, the star formation ceased shortly after the simulations started. In the result, the GC sunk quickly after that because there were no SNe. It is not clear at the moment whether the inefficient star formation indicates an improper numerical treatment of the gas physics in PoR or a failure of the QUMOND gravity.

We then explored a few additional situations less thoroughly. We ran a few Newtonian simulations with cuspy NWF halos. Even in them, without star formation, the GC can sink due to dynamical friction in a few gigayears. Star formation can but does not have to prevent the GC from sinking. It depends on the initial position and velocity of the GC. The star formation rate was reasonable with the default baryonic parameters of the code.

We then explored how our conclusions change if we change the mass of the GC. In the MOND simulations, if it is reduced to $10^4\,M_\sun$, then it often does not have difficulties surviving, even without star formation. When we increased the mass of the GC to $4\times10^5\,M_\sun$ or $10^6\,M_\sun$, which are the largest observed GC masses, the situation became the opposite. The SNe were too weak to prevent the GC from sinking, regardless of the initial tangential velocity of the GC. More precisely, {after getting close to the galaxy center,} the GC attracted and captured the central density peak of the galaxy. In the result, the galaxy would be classified as having an offset center -- or lopsided outskirts -- and the GC would be classified as a nuclear star cluster. When several massive GCs were put in the simulation, they all ended in the offset central density peak of the galaxy. A single simulation demonstrated that the sinking of massive GCs is by far not as fierce a problem in Newtonian simulations as it is in the MOND ones. 

The simulations we have performed therefore favor Newtonian gravity with NFW dark matter halos over the QUMOND gravity. However, the test is not entirely conclusive because hydrodynamics, star formation, and SN feedback are notoriously difficult to model. We identified two potential problems in PoR: 1) The code does not take into account the fact that there can be subgrid compact gas clouds that enhance gas cooling and thus star formation, and 2) PoR currently does not change the prescription for star formation from the Newtonian gravity to QUMOND gravity. While this choice was motivated by the results of \citet{Zoonozi_2021}, namely, that the free fall times of gas clouds are almost equal in the two types of gravity \citep{Nagesh_2023}, we note that \citet{Zoonozi_2021} derived their results for gas clouds of specific masses and sizes and not for a general computational cell. Furthermore, it is important to stress that in this work, we have investigated only the QUMOND version of MOND. It is not possible to conclude that our simulations disfavor MOND as a paradigm because the results can be different in other MOND theories. Indeed, MOND theories differing substantially from QUMOND have been discussed recently \citep{milg22,trimond,gqumond}. It will be interesting to see whether improvements in feedback modeling will eventually be able to at least rule out QUMOND.
 Another caveat is that our simulations have not been able to reproduce the core
that is 
often observed in isolated gas-rich dwarfs \citep{swaters09}. In such galaxies, their rotation curves suggest, assuming Newtonian gravity, cored dark matter halos with a fraction of a kiloparsec in size
\citep{Carignan1989, Gentile2007, Elson2010}. This should be able to stall the GC at 0.3-0.5 kpc in both the CDM and MOND cases. {Taken together, further research is necessary before adopting the CDM hypothesis since multiple phenomena, different from those investigated here, raise serious doubts about it \citep{kroupacjp,banik22,kroupa23}.}

\begin{acknowledgements}
We thank Joki Rosdahl for helpful advice.
This work was granted access to the HPC resources of MesoPSL financed
by the Region Ile de France and the project Equip\@Meso (reference
ANR-10-EQPX-29-01) of the programme Investissements d’Avenir supervised
by the Agence Nationale pour la Recherche. FC acknowledges support from the 
Programme National de Galaxies et Cosmologie (PNCG), du CNRS/INSU.
\end{acknowledgements}

\bibliographystyle{aa}
\bibliography{literature}

\begin{thebibliography}{87}
\expandafter\ifx\csname natexlab\endcsname\relax\def\natexlab#1{#1}\fi

\bibitem[{{Banik} {et~al.}(2022{\natexlab{a}}){Banik}, {Nagesh}, {Haghi},
  {Kroupa}, \& {Zhao}}]{Banik_fake_inclination_2022}
{Banik}, I., {Nagesh}, S.~T., {Haghi}, H., {Kroupa}, P., \& {Zhao}, H.
  2022{\natexlab{a}}, MNRAS, 513, 3541

\bibitem[{{Banik} {et~al.}(2018){Banik}, {O'Ryan}, \& {Zhao}}]{banik18}
{Banik}, I., {O'Ryan}, D., \& {Zhao}, H. 2018, \mnras, 477, 4768

\bibitem[{{Banik} {et~al.}(2020){Banik}, {Thies}, {Candlish}, {Famaey},
  {Ibata}, \& {Kroupa}}]{Banik_2020_M33}
{Banik}, I., {Thies}, I., {Candlish}, G., {et~al.} 2020, ApJ, 905, 135

\bibitem[{{Banik} {et~al.}(2022{\natexlab{b}}){Banik}, {Thies}, {Truelove},
  {Candlish}, {Famaey}, {Pawlowski}, {Ibata}, \&
  {Kroupa}}]{Banik_2022_satellite_plane}
{Banik}, I., {Thies}, I., {Truelove}, R., {et~al.} 2022{\natexlab{b}}, MNRAS,
  513, 129

\bibitem[{{Banik} \& {Zhao}(2022)}]{banik22}
{Banik}, I. \& {Zhao}, H. 2022, Symmetry, 14, 1331

\bibitem[{{Behroozi} {et~al.}(2013){Behroozi}, {Wechsler}, \&
  {Conroy}}]{behroozi13}
{Behroozi}, P.~S., {Wechsler}, R.~H., \& {Conroy}, C. 2013, \apj, 770, 57

\bibitem[{{B{\'\i}lek} {et~al.}(2022){B{\'\i}lek}, {Fensch}, {Ebrov{\'a}},
  {Nagesh}, {Famaey}, {Duc}, \& {Kroupa}}]{Bilek_2022b}
{B{\'\i}lek}, M., {Fensch}, J., {Ebrov{\'a}}, I., {et~al.} 2022, A\&A, 660, A28

\bibitem[{{B{\'\i}lek} {et~al.}(2019){B{\'\i}lek}, {Samurovi{\'c}}, \&
  {Renaud}}]{bil19b}
{B{\'\i}lek}, M., {Samurovi{\'c}}, S., \& {Renaud}, F. 2019, \aap, 629, L5

\bibitem[{{B{\'{\i}}lek} {et~al.}(2018){B{\'{\i}}lek}, {Thies}, {Kroupa}, \&
  {Famaey}}]{bil18}
{B{\'{\i}}lek}, M., {Thies}, I., {Kroupa}, P., \& {Famaey}, B. 2018, \aap, 614,
  A59

\bibitem[{{B{\'\i}lek} {et~al.}(2021){B{\'\i}lek}, {Zhao}, {Famaey},
  {M{\"u}ller}, {Kroupa}, \& {Ibata}}]{bil21}
{B{\'\i}lek}, M., {Zhao}, H., {Famaey}, B., {et~al.} 2021, \aap, 653, A170

\bibitem[{{Binney} \& {Merrifield}(1998)}]{binneymerrifield98}
{Binney}, J. \& {Merrifield}, M. 1998, {Galactic Astronomy}

\bibitem[{{Carignan} \& {Beaulieu}(1989)}]{Carignan1989}
{Carignan}, C. \& {Beaulieu}, S. 1989, \apj, 347, 760

\bibitem[{{Carleton} {et~al.}(2019){Carleton}, {Errani}, {Cooper},
  {Kaplinghat}, {Pe{\~n}arrubia}, \& {Guo}}]{Carleton2019}
{Carleton}, T., {Errani}, R., {Cooper}, M., {et~al.} 2019, \mnras, 485, 382

\bibitem[{{Chae} {et~al.}(2018){Chae}, {Bernardi}, \& {Sheth}}]{Chae_2018}
{Chae}, K.-H., {Bernardi}, M., \& {Sheth}, R.~K. 2018, ApJ, 860, 81

\bibitem[{{Ciotti} \& {Binney}(2004)}]{ciotti04}
{Ciotti}, L. \& {Binney}, J. 2004, \mnras, 351, 285

\bibitem[{{Combes}(2014)}]{combes14}
{Combes}, F. 2014, \aap, 571, A82

\bibitem[{{Di Cintio} {et~al.}(2017){Di Cintio}, {Brook}, {Dutton},
  {Macci{\`o}}, {Obreja}, \& {Dekel}}]{DiCintio2017}
{Di Cintio}, A., {Brook}, C.~B., {Dutton}, A.~A., {et~al.} 2017, \mnras, 466,
  L1

\bibitem[{{Diemer} \& {Kravtsov}(2015)}]{diemer15}
{Diemer}, B. \& {Kravtsov}, A.~V. 2015, \apj, 799, 108

\bibitem[{{Dubois} \& {Teyssier}(2008)}]{Dubois_2008}
{Dubois}, Y. \& {Teyssier}, R. 2008, A\&A, 477, 79

\bibitem[{{Eappen} \& {Kroupa}(2024)}]{eappen24}
{Eappen}, R. \& {Kroupa}, P. 2024, \mnras, 528, 4264

\bibitem[{{Eappen} {et~al.}(2022){Eappen}, {Kroupa}, {Wittenburg}, {Haslbauer},
  \& {Famaey}}]{Eappen_2022}
{Eappen}, R., {Kroupa}, P., {Wittenburg}, N., {Haslbauer}, M., \& {Famaey}, B.
  2022, MNRAS, 516, 1081

\bibitem[{{Elmegreen} \& {Falgarone}(1996)}]{Elmegreen1996}
{Elmegreen}, B.~G. \& {Falgarone}, E. 1996, \apj, 471, 816

\bibitem[{{Elson} {et~al.}(2010){Elson}, {de Blok}, \&
  {Kraan-Korteweg}}]{Elson2010}
{Elson}, E.~C., {de Blok}, W.~J.~G., \& {Kraan-Korteweg}, R.~C. 2010, \mnras,
  404, 2061

\bibitem[{{Famaey} \& {Binney}(2005)}]{famaey05}
{Famaey}, B. \& {Binney}, J. 2005, \mnras, 363, 603

\bibitem[{{Famaey} \& {McGaugh}(2012)}]{famaey12}
{Famaey}, B. \& {McGaugh}, S.~S. 2012, Living Reviews in Relativity, 15, 10

\bibitem[{{Freundlich} {et~al.}(2022){Freundlich}, {Famaey}, {Oria},
  {B{\'\i}lek}, {M{\"u}ller}, \& {Ibata}}]{freundlich22}
{Freundlich}, J., {Famaey}, B., {Oria}, P.-A., {et~al.} 2022, \aap, 658, A26

\bibitem[{{Gentile} {et~al.}(2011){Gentile}, {Famaey}, \& {de
  Blok}}]{gentile11}
{Gentile}, G., {Famaey}, B., \& {de Blok}, W.~J.~G. 2011, \aap, 527, A76

\bibitem[{{Gentile} {et~al.}(2007){Gentile}, {Salucci}, {Klein}, \&
  {Granato}}]{Gentile2007}
{Gentile}, G., {Salucci}, P., {Klein}, U., \& {Granato}, G.~L. 2007, \mnras,
  375, 199

\bibitem[{{Georgiev} {et~al.}(2009){Georgiev}, {Puzia}, {Hilker}, \&
  {Goudfrooij}}]{georgiev09}
{Georgiev}, I.~Y., {Puzia}, T.~H., {Hilker}, M., \& {Goudfrooij}, P. 2009,
  \mnras, 392, 879

\bibitem[{{Hernandez} \& {Gilmore}(1998)}]{hernandez98}
{Hernandez}, X. \& {Gilmore}, G. 1998, \mnras, 297, 517

\bibitem[{{Hui} {et~al.}(2017){Hui}, {Ostriker}, {Tremaine}, \&
  {Witten}}]{Hui2017}
{Hui}, L., {Ostriker}, J.~P., {Tremaine}, S., \& {Witten}, E. 2017, \prd, 95,
  043541

\bibitem[{{Hunter} {et~al.}(2021){Hunter}, {Elmegreen}, {Goldberger}, {Taylor},
  {Ermakov}, {Herrmann}, {Oh}, {Malko}, {Barandi}, \& {Jundt}}]{hunter21}
{Hunter}, D.~A., {Elmegreen}, B.~G., {Goldberger}, E., {et~al.} 2021, \aj, 161,
  71

\bibitem[{{Hunter} {et~al.}(2012){Hunter}, {Ficut-Vicas}, {Ashley}, {Brinks},
  {Cigan}, {Elmegreen}, {Heesen}, {Herrmann}, {Johnson}, {Oh}, {Rupen},
  {Schruba}, {Simpson}, {Walter}, {Westpfahl}, {Young}, \& {Zhang}}]{hunter12}
{Hunter}, D.~A., {Ficut-Vicas}, D., {Ashley}, T., {et~al.} 2012, \aj, 144, 134

\bibitem[{{Iocco} {et~al.}(2015){Iocco}, {Pato}, \& {Bertone}}]{iocco15}
{Iocco}, F., {Pato}, M., \& {Bertone}, G. 2015, \prd, 92, 084046

\bibitem[{{Jones} {et~al.}(2023){Jones}, {Karunakaran}, {Bennet}, {Sand},
  {Spekkens}, {Mutlu-Pakdil}, {Crnojevi{\'c}}, {Janowiecki}, {Leisman}, \&
  {Fielder}}]{Jones2023}
{Jones}, M.~G., {Karunakaran}, A., {Bennet}, P., {et~al.} 2023, \apjl, 942, L5

\bibitem[{{Karachentsev} \& {Kaisina}(2019)}]{karachentsev19}
{Karachentsev}, I.~D. \& {Kaisina}, E.~I. 2019, Astrophysical Bulletin, 74, 111

\bibitem[{{Katz}(1992)}]{katz92}
{Katz}, N. 1992, \apj, 391, 502

\bibitem[{{Kennicutt}(1989)}]{kennicutt89}
{Kennicutt}, Robert~C., J. 1989, \apj, 344, 685

\bibitem[{{Kroupa}(2015)}]{kroupacjp}
{Kroupa}, P. 2015, Canadian Journal of Physics, 93, 169

\bibitem[{{Kroupa} {et~al.}(2023){Kroupa}, {Gjergo}, {Asencio}, {Haslbauer},
  {Pflamm-Altenburg}, {Wittenburg}, {Samaras}, {Thies}, \& {Oehm}}]{kroupa23}
{Kroupa}, P., {Gjergo}, E., {Asencio}, E., {et~al.} 2023, arXiv e-prints,
  arXiv:2309.11552

\bibitem[{{L{\"u}ghausen} {et~al.}(2015){L{\"u}ghausen}, {Famaey}, \&
  {Kroupa}}]{Lughausen_2015}
{L{\"u}ghausen}, F., {Famaey}, B., \& {Kroupa}, P. 2015, Canadian Journal of
  Physics, 93, 232

\bibitem[{{L{\"u}ghausen} {et~al.}(2013){L{\"u}ghausen}, {Famaey}, {Kroupa},
  {Angus}, {Combes}, {Gentile}, {Tiret}, \& {Zhao}}]{lughausen13}
{L{\"u}ghausen}, F., {Famaey}, B., {Kroupa}, P., {et~al.} 2013, ArXiv e-prints
  [\eprint[arXiv]{1304.4931}]

\bibitem[{{Mancera Pi{\~n}a} {et~al.}(2019){Mancera Pi{\~n}a}, {Fraternali},
  {Adams}, {Marasco}, {Oosterloo}, {Oman}, {Leisman}, {di Teodoro}, {Posti},
  {Battipaglia}, {Cannon}, {Gault}, {Haynes}, {Janowiecki}, {McAllan}, {Pagel},
  {Reiter}, {Rhode}, {Salzer}, \& {Smith}}]{Mancera-Pina2019}
{Mancera Pi{\~n}a}, P.~E., {Fraternali}, F., {Adams}, E. A.~K., {et~al.} 2019,
  \apjl, 883, L33

\bibitem[{{Massari} {et~al.}(2023){Massari}, {Aguado-Agelet}, {Monelli},
  {Cassisi}, {Pancino}, {Saracino}, {Gallart}, {Ruiz-Lara},
  {Fern{\'a}ndez-Alvar}, {Surot}, {Stokholm}, {Salaris}, {Miglio}, \&
  {Ceccarelli}}]{Massari2023}
{Massari}, D., {Aguado-Agelet}, F., {Monelli}, M., {et~al.} 2023, \aap, 680,
  A20

\bibitem[{{Massey} \& {Hunter}(1998)}]{Massey1998}
{Massey}, P. \& {Hunter}, D.~A. 1998, \apj, 493, 180

\bibitem[{{Miki} \& {Umemura}(2018)}]{magi}
{Miki}, Y. \& {Umemura}, M. 2018, \mnras, 475, 2269

\bibitem[{{Milgrom}(1983{\natexlab{a}})}]{milg83b}
{Milgrom}, M. 1983{\natexlab{a}}, \apj, 270, 371

\bibitem[{{Milgrom}(1983{\natexlab{b}})}]{milg83c}
{Milgrom}, M. 1983{\natexlab{b}}, \apj, 270, 384

\bibitem[{{Milgrom}(1983{\natexlab{c}})}]{milg83a}
{Milgrom}, M. 1983{\natexlab{c}}, \apj, 270, 365

\bibitem[{{Milgrom}(2010)}]{qumond}
{Milgrom}, M. 2010, \mnras, 403, 886

\bibitem[{{Milgrom}(2022)}]{milg22}
{Milgrom}, M. 2022, \prd, 106, 064060

\bibitem[{{Milgrom}(2023{\natexlab{a}})}]{gqumond}
{Milgrom}, M. 2023{\natexlab{a}}, \prd, 108, 084005

\bibitem[{{Milgrom}(2023{\natexlab{b}})}]{trimond}
{Milgrom}, M. 2023{\natexlab{b}}, \prd, 108, 063009

\bibitem[{{Nagesh} {et~al.}(2021){Nagesh}, {Banik}, {Thies}, {Kroupa},
  {Famaey}, {Wittenburg}, {Parziale}, \& {Haslbauer}}]{Nagesh_2021}
{Nagesh}, S.~T., {Banik}, I., {Thies}, I., {et~al.} 2021, Canadian Journal of
  Physics, 99, 607

\bibitem[{{Nagesh} {et~al.}(2023){Nagesh}, {Kroupa}, {Banik}, {Famaey},
  {Ghafourian}, {Roshan}, {Thies}, {Zhao}, \& {Wittenburg}}]{Nagesh_2023}
{Nagesh}, S.~T., {Kroupa}, P., {Banik}, I., {et~al.} 2023, \mnras, 519, 5128

\bibitem[{{Navarro} {et~al.}(1996){Navarro}, {Frenk}, \& {White}}]{nfw}
{Navarro}, J.~F., {Frenk}, C.~S., \& {White}, S.~D.~M. 1996, \apj, 462, 563

\bibitem[{{Nedkova} {et~al.}(2021){Nedkova}, {H{\"a}u{\ss}ler}, {Marchesini},
  {Dimauro}, {Brammer}, {Eigenthaler}, {Feinstein}, {Ferguson},
  {Huertas-Company}, {Johnston}, {Kado-Fong}, {Kartaltepe}, {Labb{\'e}},
  {Lange-Vagle}, {Martis}, {McGrath}, {Muzzin}, {Oesch}, {Ordenes-Brice{\~n}o},
  {Puzia}, {Shipley}, {Simmons}, {Skelton}, {Stefanon}, {van der Wel}, \&
  {Whitaker}}]{nedkova21}
{Nedkova}, K.~V., {H{\"a}u{\ss}ler}, B., {Marchesini}, D., {et~al.} 2021,
  \mnras, 506, 928

\bibitem[{{Nipoti} {et~al.}(2008){Nipoti}, {Ciotti}, {Binney}, \&
  {Londrillo}}]{nipoti08}
{Nipoti}, C., {Ciotti}, L., {Binney}, J., \& {Londrillo}, P. 2008, \mnras, 386,
  2194

\bibitem[{{Pace} {et~al.}(2021){Pace}, {Walker}, {Koposov}, {Caldwell},
  {Mateo}, {Olszewski}, {Bailey}, \& {Wang}}]{Pace2021}
{Pace}, A.~B., {Walker}, M.~G., {Koposov}, S.~E., {et~al.} 2021, \apj, 923, 77

\bibitem[{{Perret} {et~al.}(2014){Perret}, {Renaud}, {Epinat}, {Amram},
  {Bournaud}, {Contini}, {Teyssier}, \& {Lambert}}]{Perret_2014}
{Perret}, V., {Renaud}, F., {Epinat}, B., {et~al.} 2014, A\&A, 562, A1

\bibitem[{{Pfenniger} \& {Combes}(1994)}]{Pfenniger1994}
{Pfenniger}, D. \& {Combes}, F. 1994, \aap, 285, 94

\bibitem[{{Planck Collaboration} {et~al.}(2016){Planck Collaboration}, {Ade},
  {Aghanim}, {Arnaud}, {Ashdown}, {Aumont}, {Baccigalupi}, {Banday},
  {Barreiro}, {Bartlett}, {Bartolo}, {Battaner}, {Battye}, {Benabed},
  {Beno{\^\i}t}, {Benoit-L{\'e}vy}, {Bernard}, {Bersanelli}, {Bielewicz},
  {Bock}, {Bonaldi}, {Bonavera}, {Bond}, {Borrill}, {Bouchet}, {Boulanger},
  {Bucher}, {Burigana}, {Butler}, {Calabrese}, {Cardoso}, {Catalano},
  {Challinor}, {Chamballu}, {Chary}, {Chiang}, {Chluba}, {Christensen},
  {Church}, {Clements}, {Colombi}, {Colombo}, {Combet}, {Coulais}, {Crill},
  {Curto}, {Cuttaia}, {Danese}, {Davies}, {Davis}, {de Bernardis}, {de Rosa},
  {de Zotti}, {Delabrouille}, {D{\'e}sert}, {Di Valentino}, {Dickinson},
  {Diego}, {Dolag}, {Dole}, {Donzelli}, {Dor{\'e}}, {Douspis}, {Ducout},
  {Dunkley}, {Dupac}, {Efstathiou}, {Elsner}, {En{\ss}lin}, {Eriksen},
  {Farhang}, {Fergusson}, {Finelli}, {Forni}, {Frailis}, {Fraisse},
  {Franceschi}, {Frejsel}, {Galeotta}, {Galli}, {Ganga}, {Gauthier}, {Gerbino},
  {Ghosh}, {Giard}, {Giraud-H{\'e}raud}, {Giusarma}, {Gjerl{\o}w},
  {Gonz{\'a}lez-Nuevo}, {G{\'o}rski}, {Gratton}, {Gregorio}, {Gruppuso},
  {Gudmundsson}, {Hamann}, {Hansen}, {Hanson}, {Harrison}, {Helou},
  {Henrot-Versill{\'e}}, {Hern{\'a}ndez-Monteagudo}, {Herranz}, {Hildebrandt},
  {Hivon}, {Hobson}, {Holmes}, {Hornstrup}, {Hovest}, {Huang}, {Huffenberger},
  {Hurier}, {Jaffe}, {Jaffe}, {Jones}, {Juvela}, {Keih{\"a}nen}, {Keskitalo},
  {Kisner}, {Kneissl}, {Knoche}, {Knox}, {Kunz}, {Kurki-Suonio}, {Lagache},
  {L{\"a}hteenm{\"a}ki}, {Lamarre}, {Lasenby}, {Lattanzi}, {Lawrence}, {Leahy},
  {Leonardi}, {Lesgourgues}, {Levrier}, {Lewis}, {Liguori}, {Lilje},
  {Linden-V{\o}rnle}, {L{\'o}pez-Caniego}, {Lubin}, {Mac{\'\i}as-P{\'e}rez},
  {Maggio}, {Maino}, {Mandolesi}, {Mangilli}, {Marchini}, {Maris}, {Martin},
  {Martinelli}, {Mart{\'\i}nez-Gonz{\'a}lez}, {Masi}, {Matarrese}, {McGehee},
  {Meinhold}, {Melchiorri}, {Melin}, {Mendes}, {Mennella}, {Migliaccio},
  {Millea}, {Mitra}, {Miville-Desch{\^e}nes}, {Moneti}, {Montier}, {Morgante},
  {Mortlock}, {Moss}, {Munshi}, {Murphy}, {Naselsky}, {Nati}, {Natoli},
  {Netterfield}, {N{\o}rgaard-Nielsen}, {Noviello}, {Novikov}, {Novikov},
  {Oxborrow}, {Paci}, {Pagano}, {Pajot}, {Paladini}, {Paoletti}, {Partridge},
  {Pasian}, {Patanchon}, {Pearson}, {Perdereau}, {Perotto}, {Perrotta},
  {Pettorino}, {Piacentini}, {Piat}, {Pierpaoli}, {Pietrobon}, {Plaszczynski},
  {Pointecouteau}, {Polenta}, {Popa}, {Pratt}, {Pr{\'e}zeau}, {Prunet},
  {Puget}, {Rachen}, {Reach}, {Rebolo}, {Reinecke}, {Remazeilles}, {Renault},
  {Renzi}, {Ristorcelli}, {Rocha}, {Rosset}, {Rossetti}, {Roudier},
  {Rouill{\'e} d'Orfeuil}, {Rowan-Robinson}, {Rubi{\~n}o-Mart{\'\i}n},
  {Rusholme}, {Said}, {Salvatelli}, {Salvati}, {Sandri}, {Santos},
  {Savelainen}, {Savini}, {Scott}, {Seiffert}, {Serra}, {Shellard}, {Spencer},
  {Spinelli}, {Stolyarov}, {Stompor}, {Sudiwala}, {Sunyaev}, {Sutton},
  {Suur-Uski}, {Sygnet}, {Tauber}, {Terenzi}, {Toffolatti}, {Tomasi},
  {Tristram}, {Trombetti}, {Tucci}, {Tuovinen}, {T{\"u}rler}, {Umana},
  {Valenziano}, {Valiviita}, {Van Tent}, {Vielva}, {Villa}, {Wade}, {Wandelt},
  {Wehus}, {White}, {White}, {Wilkinson}, {Yvon}, {Zacchei}, \&
  {Zonca}}]{Planck2016}
{Planck Collaboration}, {Ade}, P.~A.~R., {Aghanim}, N., {et~al.} 2016, \aap,
  594, A13

\bibitem[{{Rasera} \& {Teyssier}(2006)}]{Rasera_2006}
{Rasera}, Y. \& {Teyssier}, R. 2006, A\&A, 445, 1

\bibitem[{{Rejkuba}(2012)}]{rejkuba12}
{Rejkuba}, M. 2012, \apss, 341, 195

\bibitem[{{Renaud} {et~al.}(2016){Renaud}, {Famaey}, \& {Kroupa}}]{renaud16}
{Renaud}, F., {Famaey}, B., \& {Kroupa}, P. 2016, \mnras, 463, 3637

\bibitem[{{Roszkowski} {et~al.}(2018){Roszkowski}, {Sessolo}, \&
  {Trojanowski}}]{Wimp2018}
{Roszkowski}, L., {Sessolo}, E.~M., \& {Trojanowski}, S. 2018, Reports on
  Progress in Physics, 81, 066201

\bibitem[{{Roychowdhury} {et~al.}(2010){Roychowdhury}, {Chengalur}, {Begum}, \&
  {Karachentsev}}]{roychowdhury10}
{Roychowdhury}, S., {Chengalur}, J.~N., {Begum}, A., \& {Karachentsev}, I.~D.
  2010, \mnras, 404, L60

\bibitem[{{Roychowdhury} {et~al.}(2013){Roychowdhury}, {Chengalur},
  {Karachentsev}, \& {Kaisina}}]{roychowdhury13}
{Roychowdhury}, S., {Chengalur}, J.~N., {Karachentsev}, I.~D., \& {Kaisina},
  E.~I. 2013, \mnras, 436, L104

\bibitem[{{Rubin} {et~al.}(1980){Rubin}, {Ford}, \& {Thonnard}}]{Rubin1980}
{Rubin}, V.~C., {Ford}, W.~K., J., \& {Thonnard}, N. 1980, \apj, 238, 471

\bibitem[{{S{\'a}nchez-Salcedo} {et~al.}(2006){S{\'a}nchez-Salcedo},
  {Reyes-Iturbide}, \& {Hernandez}}]{sanchezsalcedo06}
{S{\'a}nchez-Salcedo}, F.~J., {Reyes-Iturbide}, J., \& {Hernandez}, X. 2006,
  \mnras, 370, 1829

\bibitem[{{Schmidt}(1959)}]{schmidt59}
{Schmidt}, M. 1959, \apj, 129, 243

\bibitem[{{Sharina} {et~al.}(2005){Sharina}, {Puzia}, \& {Makarov}}]{sharina05}
{Sharina}, M.~E., {Puzia}, T.~H., \& {Makarov}, D.~I. 2005, \aap, 442, 85

\bibitem[{{Shi} {et~al.}(2011){Shi}, {Helou}, {Yan}, {Armus}, {Wu}, {Papovich},
  \& {Stierwalt}}]{shi11}
{Shi}, Y., {Helou}, G., {Yan}, L., {et~al.} 2011, \apj, 733, 87

\bibitem[{{Skordis} \& {Z{\l}o{\'s}nik}(2021)}]{Skordis2021}
{Skordis}, C. \& {Z{\l}o{\'s}nik}, T. 2021, \prl, 127, 161302

\bibitem[{{Spitler} \& {Forbes}(2009)}]{spitler09}
{Spitler}, L.~R. \& {Forbes}, D.~A. 2009, \mnras, 392, L1

\bibitem[{{Swaters} {et~al.}(2009){Swaters}, {Sancisi}, {van Albada}, \& {van
  der Hulst}}]{swaters09}
{Swaters}, R.~A., {Sancisi}, R., {van Albada}, T.~S., \& {van der Hulst}, J.~M.
  2009, \aap, 493, 871

\bibitem[{{Tacconi} {et~al.}(2010){Tacconi}, {Genzel}, {Neri}, {Cox}, {Cooper},
  {Shapiro}, {Bolatto}, {Bouch{\'e}}, {Bournaud}, {Burkert}, {Combes},
  {Comerford}, {Davis}, {F{\"o}rster Schreiber}, {Garcia-Burillo},
  {Gracia-Carpio}, {Lutz}, {Naab}, {Omont}, {Shapley}, {Sternberg}, \&
  {Weiner}}]{Tacconi2010}
{Tacconi}, L.~J., {Genzel}, R., {Neri}, R., {et~al.} 2010, \nat, 463, 781

\bibitem[{{Teyssier}(2002)}]{Teyssier_2002}
{Teyssier}, R. 2002, A\&A, 385, 337

\bibitem[{{Teyssier} {et~al.}(2010){Teyssier}, {Chapon}, \&
  {Bournaud}}]{teyssier10}
{Teyssier}, R., {Chapon}, D., \& {Bournaud}, F. 2010, \apjl, 720, L149

\bibitem[{{Teyssier} {et~al.}(2013){Teyssier}, {Pontzen}, {Dubois}, \&
  {Read}}]{teyssier13}
{Teyssier}, R., {Pontzen}, A., {Dubois}, Y., \& {Read}, J.~I. 2013, \mnras,
  429, 3068

\bibitem[{{Thomas} {et~al.}(2017){Thomas}, {Famaey}, {Ibata}, {L{\"u}ghausen},
  \& {Kroupa}}]{thomas17}
{Thomas}, G.~F., {Famaey}, B., {Ibata}, R., {L{\"u}ghausen}, F., \& {Kroupa},
  P. 2017, \aap, 603, A65

\bibitem[{{Tiret} \& {Combes}(2007)}]{tiret07}
{Tiret}, O. \& {Combes}, F. 2007, \aap, 464, 517

\bibitem[{{Tiret} \& {Combes}(2008)}]{tiret08}
{Tiret}, O. \& {Combes}, F. 2008, in Astronomical Society of the Pacific
  Conference Series, Vol. 396, Formation and Evolution of Galaxy Disks, ed.
  J.~G. {Funes} \& E.~M. {Corsini}, 259

\bibitem[{{Wittenburg} {et~al.}(2023){Wittenburg}, {Kroupa}, {Banik},
  {Candlish}, \& {Samaras}}]{Wittenburg_2023}
{Wittenburg}, N., {Kroupa}, P., {Banik}, I., {Candlish}, G., \& {Samaras}, N.
  2023, \mnras, 523, 453

\bibitem[{{Wittenburg} {et~al.}(2020){Wittenburg}, {Kroupa}, \&
  {Famaey}}]{wittenburg20}
{Wittenburg}, N., {Kroupa}, P., \& {Famaey}, B. 2020, \apj, 890, 173

\bibitem[{{Zonoozi} {et~al.}(2021){Zonoozi}, {Lieberz}, {Banik}, {Haghi}, \&
  {Kroupa}}]{Zoonozi_2021}
{Zonoozi}, A.~H., {Lieberz}, P., {Banik}, I., {Haghi}, H., \& {Kroupa}, P.
  2021, MNRAS, 506, 5468

\bibitem[{{Zwicky}(1937)}]{Zwicky1937}
{Zwicky}, F. 1937, \apj, 86, 217

\end{thebibliography}

\begin{appendix}

\onecolumn
\vspace{-8ex}
\section{{Additional figures}}
\begin{figure*}[h] \vspace{-4ex}
        \centering 
        \includegraphics[width=17cm]{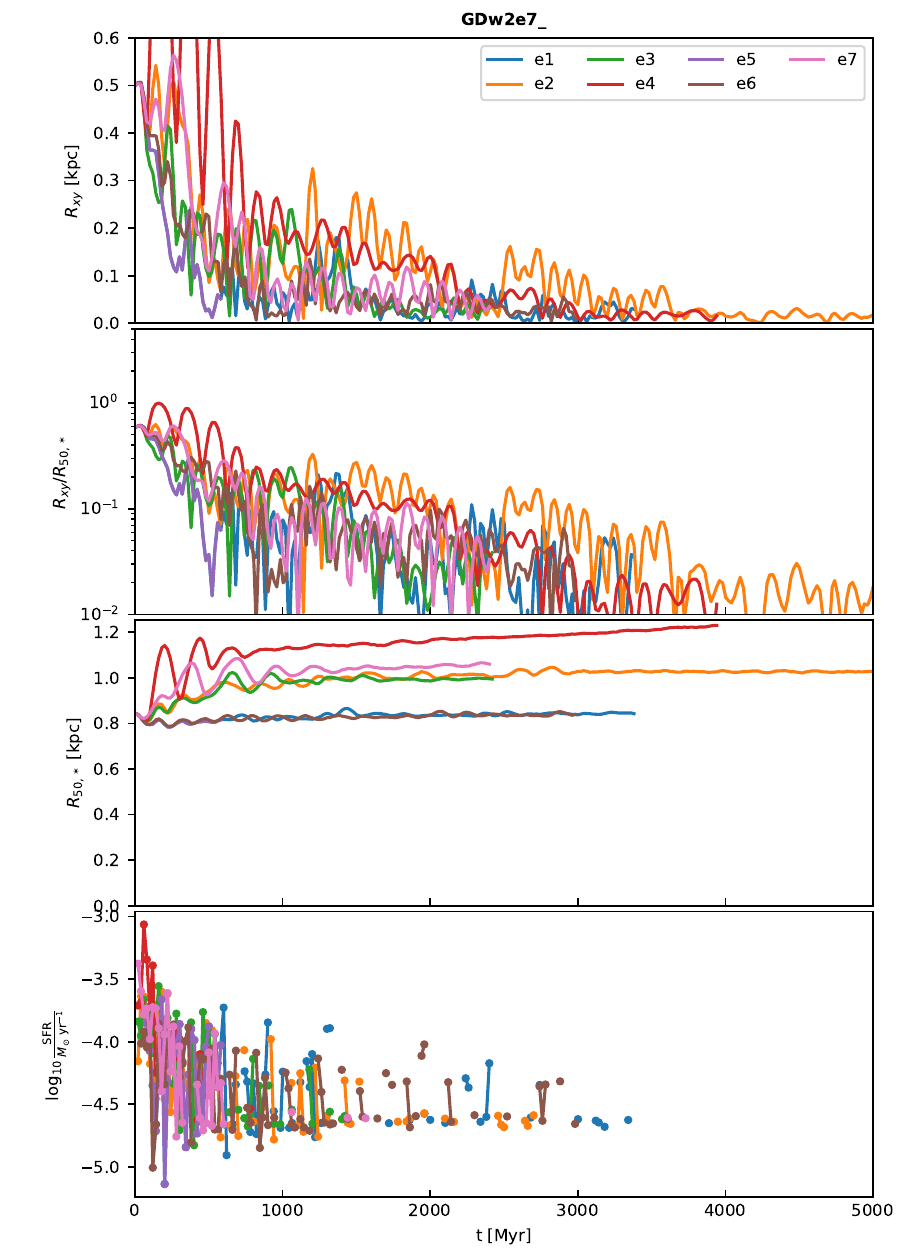}
     \caption{Different star formation variants of the model GDw2e7. (See \fig{gdw1e8} for a description.)\vspace{-8ex}} 
        \label{fig:gdw2e7}
\end{figure*}

\begin{figure*}[]
        \centering
        \includegraphics[width=17cm]{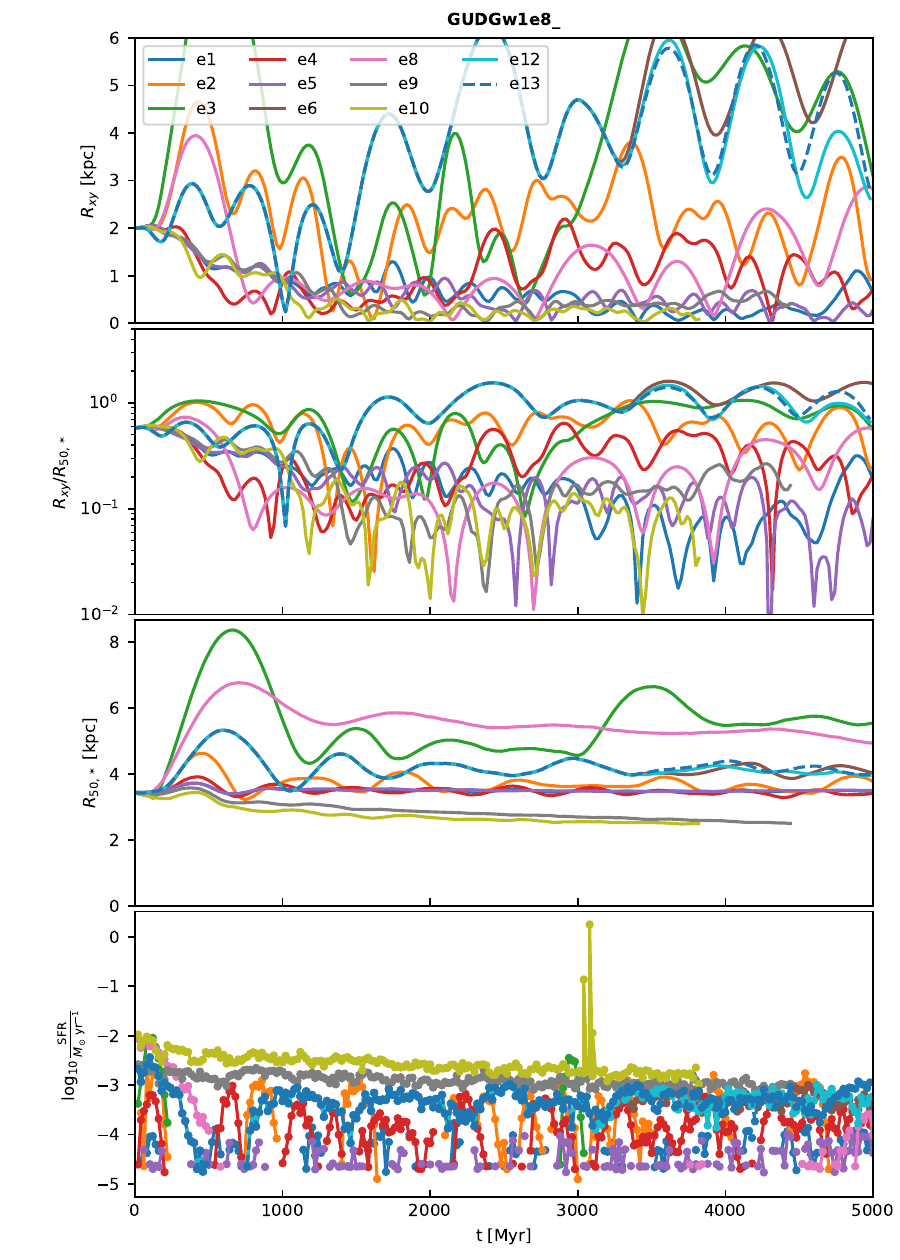}
     \caption{Different star formation variants of the model GUDG1e8. (See \fig{gdw1e8} for a description.)} 
        \label{fig:gudg1e8}
\end{figure*}

\begin{figure*}[]
        \centering
        \includegraphics[width=17cm]{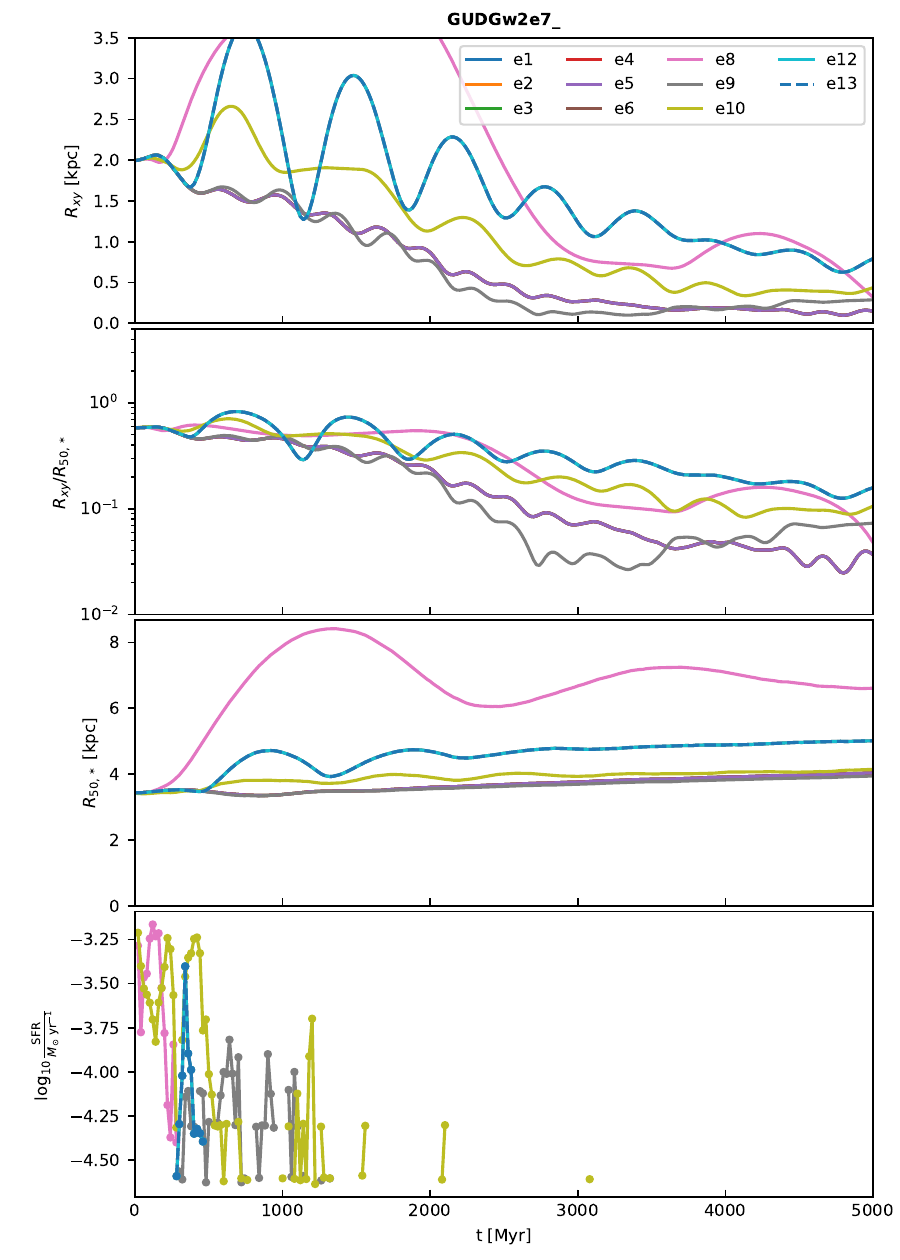}
     \caption{Different star formation variants of the model GUDG2e7. (See \fig{gdw1e8} for a description.)} 
        \label{fig:gudg2e7}
\end{figure*}
\FloatBarrier

\twocolumn

\section{Distances to observed galaxies}
\label{app:distances}

\begin{table}[htb!]
        \caption{Distances to the observed galaxies according to different sources.}
        \label{tab:distances}
        \begin{tabular}{p{1.9cm}p{1.9cm}p{1cm}p{1.1cm}p{0.8cm}}
                \hline\hline
 Name in LV & \mbox{Name in} source & Distance \mbox{in LV} [Mpc] & Distance in source [Mpc] & Source\\\hline
 \hline
KKR25 & KKR25 & 1.91 & 1.86 & S \\
UGC08508 & U8508 & 2.67 & 2.56 & S \\
DDO190 & DDO190 & 2.83 & 2.79 & S \\
ESO379-007 & E379-07 & 5.45 & 5.22 & S \\
ESO321-014 & E321-014 & 3.33 & 3.19 & S \\
KKH5 & KKH5 & 5.42 & 4.26 & S \\
KKH34 & KKH34 & 7.28 & 4.61 & S \\
KKH98 & KKH98 & 2.58 & 2.45 & S \\
KK16 & KK16 & 5.62 & 4.74 & S \\
KK17 & KK17 & 5.01 & 4.72 & S \\
KKH18 & KKH18 & 4.79 & 4.43 & S \\
ESO489-056 & E489-56 & 6.31 & 4.99 & S \\
ESO490-017 & E490-17 & 6.34 & 4.23 & S \\
UGC03755 & U3755 & 7.69 & 5.22 & S \\
KK65 & KK65 & 7.98 & 4.51 & S \\
UGC04115 & U4115 & 7.87 & 5.49 & S \\
KKH86 & KKH86 & 2.61 & 2.61 & S \\
UGCA438 & UA438 & 2.22 & 2.23 & S \\
LSBCD634-03 & D634-03 & 9.59 & 9.46 & G \\
DDO052 & DDO52 & 9.86 & 10.28 & G \\
ESO121-020 & ESO121-20 & 6.08 & 6.05 & G \\
HIPASSJ1247-77 & HIPASSJ1247-77 & 3.47 & 3.16 & G \\
HS117 & HS117 & 3.96 & 3.96 & G \\
IC4662 & IC4662 & 2.55 & 2.44 & G \\
KK182 & KK182 & 5.94 & 5.78 & G \\
KK230 & KK230 & 2.21 & 1.92 & G \\
KK246 & KK246 & 6.85 & 7.83 & G \\
UGC07242 & KKH77 & 5.45 & 5.42 & G \\
NGC4605 & NGC4605 & 5.55 & 5.47 & G \\
IC779 & UGC7369 & 16.67 & 11.60 & G \\

                \hline
        \end{tabular}
        \tablefoot{{``LV'' stands for the Local Volume database \citep{karachentsev19} (\sect{obs}). ``Source'' refers to the publication from which we took the parameters of the GCs: ``S'' stands for \citet{sharina05} and ``G'' for \citet{georgiev09}.} }
\end{table}

\vfill\eject

\section{Stability of the simulated NFW halo}
\label{app:nfwstab}
\begin{figure}[!hbt]
        \resizebox{\hsize}{!}{\includegraphics{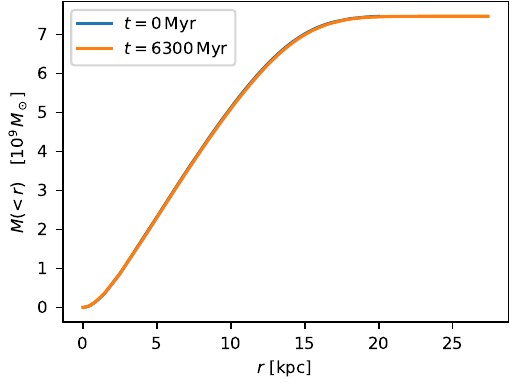}}
        \caption{Demonstration that the MAGI code generates stable dark matter halos. Blue curve: the cumulative mass profile of the dark matter particles at the beginning of the simulation GDw1e8N without star formation. Orange curve: the same profile at the end of the simulation, 6300\,Myr later. } 
        \label{fig:nfwstab}
\end{figure}

\FloatBarrier

\section{Star formation parameters}
\label{app:parameters}

\begin{table}[!htb]
 \centering
\caption{Variants of the model GDw1e8. }
\label{tab:vardw1e8}
\begin{tabular}{lllll}
\hline\hline
Variant  & \url{n_star}  & \url{t_star} & \url{eta_sn} & \url{f_ek}  \\\hline 
\url{GDw1e8_e1} &  0.5  & 3.0 & 0.1 & 0.5   \\
\url{GDw1e8_e2} &  0.1  & 3.0 & 0.1 & 0.5   \\
\url{GDw1e8_e3} &  0.1  & 6.0 & 0.1 & 0.5   \\
\url{GDw1e8_e4} &  0.01 & 3.0 & 0.1 & 0.5   \\
\url{GDw1e8_e6} &  2    & 3.0 & 0.1 & 0.5   \\
\url{GDw1e8_e7} &  0.1  & 10.0 & 0.1 & 0.5   \\
\hline
\end{tabular}
\end{table}

\begin{table}[h!]
 \centering
\caption{Variants of the model GDw2e7. }
\label{tab:vardw2e7}
\begin{tabular}{lllll}
\hline\hline
Variant   & \url{n_star}  & \url{t_star} & \url{eta_sn} & \url{f_ek} \\\hline 
\url{GDw2e7_e1}   & 0.5  & 3.0 & 0.1 & 0.5 \\
\url{GDw2e7_e2}   & 0.1  & 3.0 & 0.1 & 0.5 \\
\url{GDw2e7_e3}   & 0.1  & 2.0 & 0.1 & 0.5 \\
\url{GDw2e7_e4}   & 0.1  & 1.0 & 0.1 & 0.5 \\
\url{GDw2e7_e5}   & 0.5  & 3.0 & 0.1 & 1    \\
\url{GDw2e7_e6}   & 0.5  & 3.0 & 0.1 & 0.5, Z=0.001 \\
\url{GDw2e7_e7}   & 0.5  & 1.0 & 0.1 & 0.5, $r_\mathrm{bubble}$=50pc  \\
\hline
\end{tabular}
\tablefoot{For the {GDw2e7\_e6} model, the default gas metallicity of 0.1 was decreased to 0.001. For the {GDw2e7\_e7} model, the SN bubble radius $r_\mathrm{bubble}$ was changed from the default value of 150\,pc to 50\,pc. }
\end{table}

\begin{table}[h!]
 \centering
\caption{Variants of the models GUDG1e8 and GUDG2e7.}
\label{tab:varudg1e8}
\begin{tabular}{lllll}
\hline\hline
Variant   & \url{n_star}  & \url{t_star} & \url{eta_sn} & \url{f_ek}  \\\hline 
\url{GUDG1e8_e1}  & 0.1  & 3.0 & 0.1 & 0.5  \\
\url{GUDG1e8_e2}  & 0.1  & 0.3 & 0.1 & 0.5   \\
\url{GUDG1e8_e3}  &  0.05  & 0.3 & 0.1 & 0.5  \\
\url{GUDG1e8_e4}  &  0.05  & 3.0 & 0.1 & 0.5  \\
\url{GUDG1e8_e5}  &  0.1  & 6.0 & 0.1 & 0.5   \\
\url{GUDG1e8_e6}  &  0.01  & 3.0 & 0.1 & 0.5   \\
\url{GUDG1e8_e8}  &  0.001   & 3.0 & 0.1 & 0.5  \\
\url{GUDG1e8_e9}  &  0.01   & 3.0 & 0.01 & 0.5   \\
\url{GUDG1e8_e10} &  0.001   & 3.0 & 0.01 & 0.5  \\
\url{GUDG1e8_e12} &  0.01  & 3.0 & 0.1 & 0.0  \\
\url{GUDG1e8_e13} &  0.01  & 3.0 & 0.1 & 1.0  \\
\hline
\end{tabular}
\tablefoot{ The parameters are given only for the GUDG1e8 models. For the GUDG2e7 models, we used the same sets of parameters. }
\end{table}

\begin{table}[h!]
 \centering
\caption{Star formation and feedback parameters of the Newtonian models.}
\label{tab:varnewt}
\begin{tabular}{llllll}
\hline\hline
Model   & \url{n_star} & \url{t_star} & \url{eta_sn} & \url{f_ek} \\\hline 
\url{GDw1e8N}  & 0.1 & 6.0 & 0.1 & 0.5 \\
\url{GDw2e7N}  & 0.1 & 3.0 & 0.1 & 0.5 \\
\hline
\end{tabular}
\tablefoot{ Only one variant was considered for each model.}
\end{table}

\end{appendix}

\end{document}